\documentclass[aps,twocolumn,floatfix,altaffilletter,superscriptaddress,preprintnumbers,tightenlines,showpacs,showkeys,notitlepage,nofootinbib]{revtex4-1}

\usepackage[T1]{fontenc}
\usepackage[colorlinks=true,citecolor=blue,linkcolor=blue,urlcolor=blue]{hyperref}
\usepackage{braket,caption,
graphicx,physics,amsthm,amssymb,amsfonts,xcolor,booktabs,siunitx,multirow,tablefootnote}
\usepackage{mathtools}
\usepackage{slashed,bm}
\usepackage[capitalise, english]{cleveref}
\usepackage[normalem]{ulem}
\usepackage{xspace}
\usepackage{mathrsfs}
\usepackage{soul}
\usepackage{fontawesome}

\usepackage{cleveref}
\usepackage{tikz}
\usetikzlibrary{arrows,shapes}
\usetikzlibrary{trees}
\usetikzlibrary{snakes}
\usetikzlibrary{matrix,arrows}
\usepackage{subfig}
\usetikzlibrary{positioning}				
\usetikzlibrary{calc,through}				
\usetikzlibrary{decorations.pathreplacing} 
\usepackage[tikz]{bclogo} 					
\usepackage{pgffor}							
\usetikzlibrary{decorations.pathmorphing}	
\usetikzlibrary{decorations.markings}
\usetikzlibrary{intersections}		

\tikzset{
    vector/.style={decorate, decoration={snake}, draw},
    fermion/.style={postaction={decorate},
        decoration={markings,mark=at position .55 with {\arrow{>}}}},
    fermionbar/.style={draw, postaction={decorate},
        decoration={markings,mark=at position .55 with {\arrow{<}}}},
    fermionnoarrow/.style={},
    gluon/.style={decorate,
        decoration={coil,amplitude=4pt, segment length=5pt}},
    scalar/.style={dashed, postaction={decorate},
        decoration={markings,mark=at position .55 with {\arrow{>}}}},
    scalarbar/.style={dashed, postaction={decorate},
        decoration={markings,mark=at position .55 with {\arrow{<}}}},
    scalarnoarrow/.style={dashed,draw},
	vectorscalar/.style={loosely dotted,draw=black, postaction={decorate}},
}

\makeatletter
\providecommand*{\diff}%
	{\@ifnextchar^{\DIfF}{\DIfF^{}}}
\def\DIfF^#1{%
	\mathop{\mathrm{\mathstrut d}}%
		\nolimits^{#1}\gobblespace}
\def\gobblespace{%
	\futurelet\diffarg\opspace}
\def\opspace{%
	\let\DiffSpace\!%
	\ifx\diffarg(%
		\let\DiffSpace\relax
	\else
		\ifx\diffarg[%
			\let\DiffSpace\relax
		\else
  			\ifx\diffarg\{%
				\let\DiffSpace\relax
			\fi\fi\fi\DiffSpace}

\definecolor{cbred}{HTML}{ff0000}
\definecolor{cborange}{HTML}{FFA500}
\definecolor{cbgreen}{HTML}{008000}
\definecolor{cbyellow}{HTML}{f1dd42}
\definecolor{cblblue}{HTML}{56b4e9}
\definecolor{cbblue}{HTML}{0072b2}
\definecolor{defgrey}{HTML}{9f9f9f}
\definecolor{defgreen}{HTML}{8eba42}
\definecolor{defgrey}{HTML}{808080}
\definecolor{ercolor}{HTML}{00FFFF}
\definecolor{nrcolor}{HTML}{FF00FF}
\definecolor{sigcolor}{HTML}{FFE600}
\definecolor{yellow35}{HTML}{ffae17}
\definecolor{green35}{HTML}{9fac43}
\definecolor{blueWIMP}{HTML}{0000aa}

\definecolor{Tsallis}{HTML}{dfa8be}
\definecolor{Empirical}{HTML}{f1cb98}
\definecolor{Boltzmann}{HTML}{9ccff1}

\definecolor{fig34red}{HTML}{99434b}
\definecolor{fig34blue}{HTML}{4169E1}

\newcommand{\de}{\mathrm{d}}

\newcommand{\ch}{\mathcal{H}}

\newcommand{\rhov}{\rho_{\rm DS}}

\newcommand{\rD}{\rho_{\rm DS}}

\newcommand{\hats}{\hat{\sigma}}
\newcommand{\Ods}{\Omega_{\rm DS}}

\newcommand{\hatr}{\hat{\rho}}
\newcommand{\hatp}{\hat{p}}

\newcommand{\be}{\begin{equation}}
\newcommand{\ee}{\end{equation}}

\def\l@subsection#1#2{}
\def\l@subsubsection#1#2{}

\begin{document}

\title{Post-Recombination Fluctuations from a Sequestered Dark Sector}
\author{Salvatore Bottaro}
\affiliation{Raymond and Beverly Sackler School of Physics and Astronomy, Tel Aviv University, Tel Aviv, Israel}
\affiliation{Department of Particle Physics and Astrophysics, Weizmann Institute of Science, Rehovot, Israel}
\author{Michael Geller}
\affiliation{Raymond and Beverly Sackler School of Physics and Astronomy, Tel Aviv University, Tel Aviv, Israel}
\author{Diego Redigolo}
\affiliation{INFN, Sezione di Firenze, Via G. Sansone 1, Sesto Fiorentino, Italy}
\author{Maya Tsur}
\affiliation{Raymond and Beverly Sackler School of Physics and Astronomy, Tel Aviv University, Tel Aviv, Israel}

\begin{abstract}
We develop a formalism to characterize the imprints of late-time sources of cosmological fluctuations under the sole assumption that the injection occurs on timescales short compared to the horizon. For post-recombination injections, we derive the general modification of photon geodesics in the presence of scalar, vector, and tensor perturbations, and compute the resulting impact on the Cosmic Microwave Background through the integrated Sachs–Wolfe effect. We show that the signal is generically dominated by instantaneous injections of anisotropic stress. As an application, we consider first-order phase transitions in a sequestered dark sector and show that current observations constrain fractional energy injections at the permille level.
\end{abstract}

\maketitle
\section{Introduction}
We constrain cosmological fluctuations sourced by dynamics in a sequestered dark sector (DS) that interacts with the Standard Model (SM) only through gravity. While observations of the Cosmic Microwave Background (CMB) indicate that primordial fluctuations are small and nearly scale invariant, additional contributions may arise throughout cosmic history from DS processes such as phase transitions, topological defects, tachyonic instabilities, or particle production~\cite{Kibble1980,
Witten1984,GuthTye1980,Kibble1976,Zurek1985,VilenkinShellard1994,
HindmarshKibble1995,FelderKofmanLinde2001PRD,FelderEtAl2001PRL,Parker1968,KofmanLindeStarobinsky1994,KofmanLindeStarobinsky1997,
MazumdarWhite2019,AminEtAl2014}. While typically associated with high-energy physics in the early Universe, such mechanisms may also arise at late times in scenarios involving feebly coupled ultra-light degrees of freedom, for instance in non-minimal models of dark energy such as quintessence~\cite{Creminelli:2008wc}.

We construct the macroscopic stress-energy tensor of a sequestered DS coupled gravitationally to the SM and dark matter (DM). In the limit where the duration of the DS process is short compared to the Hubble time at the transition, the resulting fluctuations are controlled by a small set of parameters: the characteristic scale relative to the Hubble horizon, $\beta_H \equiv \beta_*/\mathcal{H}_*$, the activation time $a_*$, and the DS energy density $\Omega_{\rm DS} \equiv \rho_{\rm DS}/(\rho_M+\rho_{\rm DS})|_{a_*}$.\footnote{We adopt conformal time $d\eta = dt/a$, for which $\mathcal{H}_* = a_* H_*$ and $\beta_* = a_* \beta$.}

In this regime, the DS contribution is fully specified by five equal-time correlators of the spatial stress-energy tensor. During matter domination, the phenomenology simplifies: a sudden variation of the gravitational potentials sources an integrated Sachs--Wolfe (ISW) signal that would otherwise vanish. We show that the dominant contribution is controlled by the most non-analytic component of the source, namely the instantaneous injection of anisotropic stress.

We compute this signal for bubble collisions in a first-order phase transition during matter domination within the envelope approximation, extending Ref.~\cite{Jinno:2016vai}. We derive strong constraints on such scenarios: for $\beta_H \sim 10$, the fractional energy density at injection is bounded to be $\lesssim 10^{-3}$, while for larger $\beta_H$ the bounds weaken rapidly, allowing $\mathcal{O}(10^{-1})$ before backreaction on the background expansion becomes significant. This delineates the observable window for late-time dark-sector phase transitions.

\section{The Setup}
\begin{figure}[hpt!]
    \centering
    \includegraphics[width=0.85\linewidth]{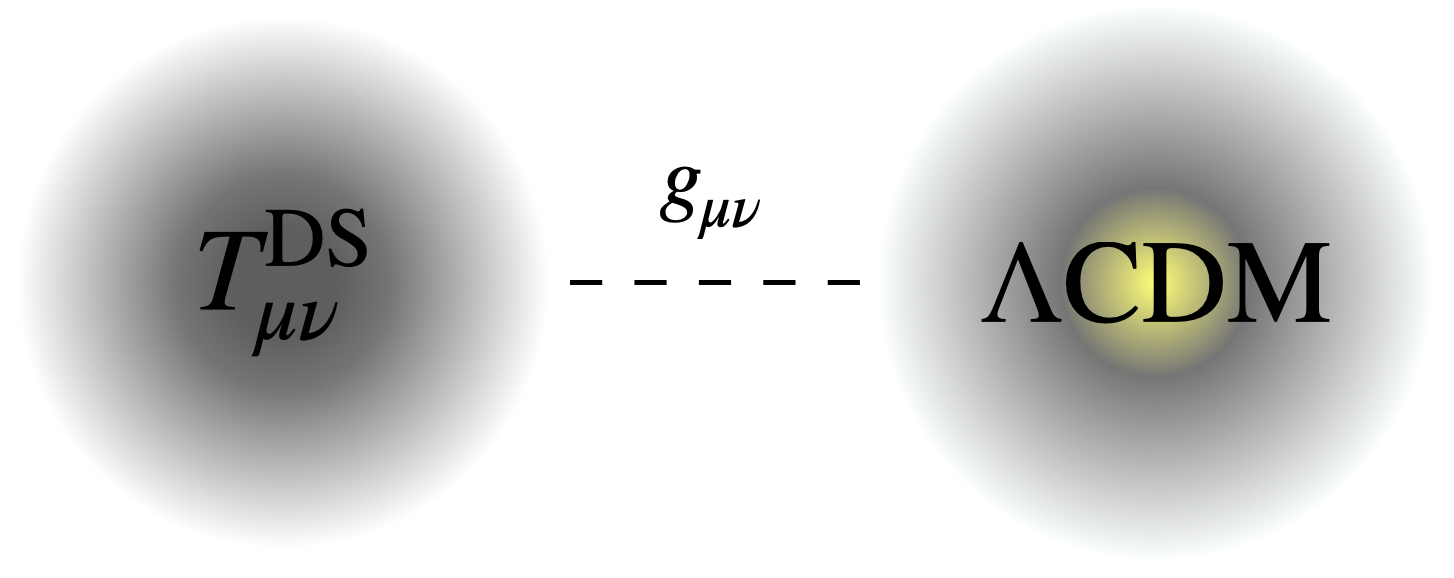}
    \caption{Cartoon of our setup where a DS which constitutes a small fraction of the Universe energy density is coupled to $\Lambda$CDM purely gravitationally.}
    \label{fig:sketch}
\end{figure}
We consider a dark sector (DS) coupled only gravitationally to $\Lambda$CDM. Its stress-energy tensor $T_{\mu\nu}^{\rm DS}$, being sequestered, is separately covariantly conserved,
\begin{equation}\label{eq:cov_cons}
    \nabla_{\mu}T_{\nu}^{\mu,\rm DS}=0\ .
\end{equation}

This condition constrains both the background $\overline{T}_{\nu}^{\mu,\rm DS}$ and perturbations $\delta T_{\nu}^{\mu,\rm DS}$. The most general background consistent with Eq.~\eqref{eq:cov_cons} together with homogeneity and isotropy is\footnote{In principle, a $\delta(\eta-\eta_*)$ contribution would be compatible with Eq.~\eqref{eq:cov_cons}. However, through the Friedmann equations it would induce a distributional feature in $\mathcal{H}$, corresponding to an infinitely thin shell in time, which lies outside the class of cosmological evolutions considered here.}  

\begin{widetext}
    \begin{equation}\label{eq:background}
\begin{split}
    &\overline{T}_{0}^{0,\rm DS}=\rD\left(\left(\frac{a_*}{a}\right)^{3(1+w_<)}\theta(\eta_*-\eta)+\left(\frac{a_*}{a}\right)^{3(1+w_>)}\theta(\eta-\eta_*)\right)\, ,\\
    &\overline{T}_{j}^{i,\rm DS}=-\delta^i_j\rD\left(w_<\left(\frac{a_*}{a}\right)^{3(1+w_<)}\theta(\eta_*-\eta)+w_>\left(\frac{a_*}{a}\right)^{3(1+w_>)}\theta(\eta-\eta_*)\right)\ ,
    \end{split}
\end{equation}
\end{widetext}

where $a \equiv a(\eta)$ and $a_* \equiv a(\eta_*)$ (during matter domination, $a\propto\eta^2$). $\rho_{\rm DS}$ is the DS energy density and $w_<$ ($w_>$) the equation of state before (after) $a_*$. For a first-order phase transition, typically $w_<=-1$ and $w_>=1/3$. 

Consistently, Eq.~\eqref{eq:cov_cons} constrains the leading singular structure of the fluctuations in the short-duration expansion, 
\begin{equation}\label{eq:disc}
\begin{split}
    & \delta T_{0}^0=\frac{\rho_{\rm DS}}{\beta_H}\hat{T}_{0}^0(\vec{x},\eta)\theta(\eta-\eta_*)\ ,\\
    &\delta T_{i}^0=\frac{\rho_{\rm DS}}{\beta_H}\hat{T}_{i}^0(\vec{x},\eta)\theta(\eta-\eta_*)\ ,\\
    & \delta T_{j}^i=\frac{\rho_{\rm DS}}{\beta_*}\hat{T}_{j}^i(\vec{x})\delta(\eta-\eta_*)\ ,
\end{split}
\end{equation}
where the continuous terms proportional to $\theta(\eta-\eta_*)$ are matched to enforce covariant conservation once the instantaneous injections proportional to $\delta(\eta-\eta_*)$ are specified. We take $\beta_*$ to set both the characteristic time and length scales of the process (assuming a relativistic source); for a phase transition, this corresponds to the comoving bubble size at percolation~\cite{Turner:1992tz}.

Using the scalar-vector-tensor decomposition we write 
$\hat{T}_{0}^0=\hat{\rho}$, $\hat{T}_{i}^0=-\partial_i \hat{v}-\hat{v}^T_{i}$, and $\hat{T}_{j}^i=-\delta_j^i\hat{p}-\left(\frac{\partial^i\partial_j}{\nabla^2}-\frac{1}{3}\delta_j^i\right)\hat{\sigma}
-\partial_{(i}\hat{\pi}^T_{j)}-\hat{\pi}_{ij}^{TT}$. In this approximation, the unequal-time correlators characterizing the source at linear order~\cite{Caprini:2009fx} reduce to equal-time correlators evaluated at $\eta_*$. These are specified by the characteristic scale $\beta_*$ and five form factors encoding the spatial dependence of the two-point correlators: three for the scalar components (pressure, anisotropic stress, and their cross-correlation), one for the vector, and one for the tensor~\cite{Ma:1995ey}. All in all we get
\begin{align}\label{eq:ff}
&\langle \hatp(\vec{k})\hatp(\vec{q})\rangle=\left(\frac{2\pi}{\beta_*}\right)^3\delta(\vec{k}+\vec{q})\mathcal{P}_{pp}(k/\beta_*)\,,\notag\\
    &\langle \hats(\vec{k})\hatp(\vec{q})\rangle=\left(\frac{2\pi}{\beta_*}\right)^3\delta(\vec{k}+\vec{q})\mathcal{P}_{p\sigma}(k/\beta_*)\,,\notag\\
    &\langle \hats(\vec{k})\hats(\vec{q})\rangle=\left(\frac{2\pi}{\beta_*}\right)^3\delta(\vec{k}+\vec{q})\mathcal{P}_{\sigma\sigma}(k/\beta_*)\,,\\
    &\langle \hat{\pi}^T_i(\vec{k})\hat{\pi}^T_j(\vec{q})\rangle=\frac{1}{k^2}\Pi_{ij}^{T}\left(\frac{2\pi}{\beta_*}\right)^3\delta(\vec{k}+\vec{q})\mathcal{P}_{T}(k/\beta_*)\,,\notag\\
    &\langle \hat{\pi}^{TT}_{ij}(\vec{k})\hat{\pi}^{TT}_{mn}(\vec{q})\rangle=\Pi_{ij,mn}^{TT}\left(\frac{2\pi}{\beta_*}\right)^3\!\!\!\delta(\vec{k}+\vec{q})\mathcal{P}_{TT}(k/\beta_*)\,,\notag
\end{align}
where we have defined the projectors $\Pi_{ij}^T=\left(\frac{k_i k_j}{k^2}-\delta_{ij}\right)$ and $\Pi_{ij,mn}^{TT}=(\Pi^T_{m(i}\Pi^T_{j)n}-\Pi_{ij}^T\Pi_{mn}^T)/2$.The other stress tensor components are fixed by covariant conservation, as detailed in Appendix~\ref{sec:fulltransfer}.

For a relativistic single-scale processes, dimensional analysis factors out the dependence on $\beta_*$ in the equal-time correlators, leaving dimensionless shape functions encoding the microscopic origin of the source. Their observable impact is mediated by the induced gravitational potentials, which subsequently affect photon geodesics and source matter perturbations.

\section{Gravitational response and Integrated Sachs-Wolfe}
By solving the linearized Einstein equations with the sources derived in Eq.~\eqref{eq:disc} we can get the gravitational fluctuations in the metric\footnote{We adopt the conventions of Ref.~\cite{Gorbunov:2011zzc}, with $\Phi \to \psi$ and $\Psi \to -\phi$ in the notation of Ref.~\cite{Ma:1995ey}.} 
\begin{equation}
\de s^2=a^2(\eta)\left((1+\delta g_{00})\de \eta^2-\left(\delta_{ij}+\delta g_{ij}\right)\de x^i \de x^j\right)\,,
\end{equation}
where $\delta g_{00}=2\Phi$ and $\delta g_{ij}=2\Psi\delta_{ij}-\partial_{(i}W_{j)}-h_{ij}$ with $\Phi$ being Newtonian potential and $\Psi$ the spatial curvature. Gravitational fluctuations affect photon geodesics through the ISW effect, which in the presence of scalar, vector, and tensor modes can be written as an integral over time and Fourier modes, where the integrand depends on the time derivatives of all gravitational perturbations. In particular, the ISW contribution shows that photon geodesics from the injection time $\eta_*$ until today are sensitive to the time derivatives of all metric fluctuations
\begin{align}
\Theta(\hat n)
&= \int \frac{\mathrm{d}^3 k}{(2\pi)^3}
   \int_{\eta_*}^{\eta_0}\mathrm{d}\eta\,
   e^{i(\eta_0-\eta)\vec{k}\cdot\hat n}\gamma(\hat{n},\vec{k})\ ,
\label{eq:ISW}
\end{align}
where $\gamma(\hat{n},\vec{k})=\left(
\Phi' - \Psi'
+ i n_i n_j k_j W'_i
-\frac12 n_i n_j h'_{ij}
\right)$, $\hat{n}$ is the photon direction and the gravitational fluctuations are functions of the conformal time and the spatial momentum. 

In general, the metric perturbations can be expressed in terms of the injected sources through appropriate transfer functions. The scalar potential receives contributions from anisotropic stress and pressure, while vector and tensor modes are sourced by transverse and transverse-traceless components of the stress tensor. These contributions are weighted by transfer functions that encode the time evolution of each mode
\begin{align}
&\Psi(\eta,k)=\frac{\Omega_{\rm{DS}}}{\beta_H}\left[ T_\sigma(\eta, k) \hat{\sigma}(\vec{k})+T_p(\eta,k) \hat{p}(\vec{k})\right] \, ,  \label{eq:scalar} \\
&W_{i}(\eta,k)=\frac{\Omega_{\rm{DS}}}{\beta_H} T_v(\eta) \hat{\pi}_i^T(\vec{k})\, ,\label{eq:vec}\\
&h_{ij}(\eta,k)= \frac{\Omega_{\rm{DS}}}{\beta_H}T_h(\eta,k) \hat{\pi}_{ij}^{TT}(\vec{k})\, .\label{eq:tensor}
\end{align}
The full transfer functions are given in Appendix~\ref{sec:fulltransfer}. Unlike scalars and tensors, the vector transfer function depends only on time, reflecting the decay of injected vector modes. A constraint relates the two scalar potentials at the time of injection, fixing a discontinuity proportional to the anisotropic stress 
\begin{equation}
 \Phi+\Psi=\frac{3\mathcal{H}_*}{k^2}\frac{\Ods}{\beta_H}\delta(\eta-\eta_*)\hat{\sigma}\ .  \label{eq:sigma} 
\end{equation}
After the injection, the two potentials coincide, and their subsequent evolution is fully captured by the scalar transfer functions in Eq.~\eqref{eq:psi}.

A general comment on the analytic structure of the metric perturbations is important. Vector and tensor modes obey second-order differential equations in time. As a result, an impulsive source in the stress tensor such as the one in Eq.~\eqref{eq:disc} produces continuous responses in these modes. In contrast, an instantaneous injection of anisotropic stress induces a delta-function discontinuity in the scalar potentials, followed by a step-like contribution at later times. An instantaneous pressure injection instead sources scalar sound modes, which are governed by second-order equations and therefore remain continuous in time. 

We can now put Eq.~\eqref{eq:ISW} and the time evolutions of the gravitational perturbation in \cref{eq:scalar,eq:vec,eq:tensor} together to rewrite the ISW contribution to the CMB power spectrum $\langle\Theta(\hat{n})\Theta(\hat{n}')\rangle=\sum_\ell\frac{2\ell+1}{2\ell(\ell+1)\bar{T}^2} \mathcal{D}_\ell P_\ell(\hat{n}\cdot\hat{n}')$ directly in terms of the five form factors defined in Eq.~\eqref{eq:ff} describing spatial behavior of the source at injection time 
\begin{widetext}
\begin{equation}\label{eq:ISWfull}
\mathcal{D}_\ell= \frac{\Ods^2}{\beta_H^5}\frac{\ell(\ell+1)\bar{T}^2}{\pi^2}\int \frac{k^2\mathrm{d}k}{\ch_*^3}\left[\sum_{i,j}^{p,\sigma}\mathcal{S}_{i,\ell}(k)\mathcal{S}_{j,\ell}(k)\mathcal{P}_{ij}\!\left(\frac{k}{\beta_*}\right)+\mathcal{V}_{\ell}^2(k)\mathcal{P}_{T}\left(\frac{k}{\beta_*}\right)+\mathcal{T}_\ell^{2}(k)\mathcal{P}_{TT}\left(\frac{k}{\beta_*}\right)\right]   ,
\end{equation}
\end{widetext}
where $\bar{T}$ is the CMB background temperature today and we defined the kernels of the different helicities as
\begin{align}\label{eq:kernels}
    &\mathcal{S}_{\sigma,\ell}=\frac{3\ch_*}{k}j_\ell'(k\chi_*)+2\int_{\eta_*}^{\eta_0}\mathrm{d}\eta j_\ell(k\chi)\partial_\eta T_\sigma(k,\eta)\, ,\notag\\
    &\mathcal{S}_{p,\ell}=2\int_{\eta_*}^{\eta_0}\mathrm{d}\eta j_\ell(k\chi)\partial_\eta T_p(k,\eta)\, ,\\
    &\mathcal{V}_\ell=\sqrt{\frac{\ell(\ell+1)}{2(2\ell+1)^2}}\!\int_{\eta_*}^{\eta_0}\!\!\!\mathrm{d}\eta\left(j_{\ell+1}'(k\chi)+j_{\ell-1}'(k\chi)\right)\partial_\eta T_v(\eta)\notag\, ,\\
    &\mathcal{T}_\ell=\sqrt{\frac{(\ell+2)!}{8(\ell-2)!}}\int_{\eta_*}^{\eta_0}\mathrm{d}\eta\frac{j_{\ell}(k\chi)}{k^2\chi^2}\partial_\eta T_h(\eta,k)\, ,\notag
\end{align}
where we defined $\chi\equiv \eta_0-\eta$, $\chi_*\equiv \eta_0-\eta_*$, while $j_\ell(x)$ denotes the $\ell^{\rm th}$ spherical Bessel function of the first kind. A few remarks are in order. First, Eq.~\eqref{eq:ISWfull} captures the ISW contribution of a generic injection under the sole assumption that $\beta_H \gg 1$, while the kernels in Eq.~\eqref{eq:kernels} encode the interplay between gravitational transfer functions and angular projection. Second, the characteristic scale of the process, $\beta_*$, maps onto a characteristic multipole $\ell_*\simeq \beta_H \left(\frac{\eta_0}{\eta_*}-1\right)$, a relation that follows directly from the Limber approximation~\cite{LoVerde:2008re}.

Finally, the ISW signal is controlled by the non-analytic structure of the gravitational potentials. At the characteristic multipole $\ell_*$, the projection kernel exhibits oscillations on a timescale $\Delta\eta \sim \eta_*/\beta_H$. The response then depends sensitively on the temporal structure of the source. Localized sources in time effectively probe the kernel at a single instant and therefore produce unsuppressed contributions. In contrast, sources that vary smoothly over a Hubble timescale sample many oscillations of the kernel—of order $\mathcal{O}(\beta_H)$—which partially cancel upon integration. As a result, their contribution is suppressed by a factor $\sim 1/\beta_H$ relative to the localized case. A detailed derivation of this scaling, based on the asymptotic behavior of spherical Bessel functions, is presented in Appendix~\ref{sec:approx_bessel}, where we also provide numerical checks confirming its validity.

An immediate consequence of this scaling is that late-time fluctuation injections are predominantly imprinted in the CMB through their anisotropic stress component. Moreover, late-time observables probing the matter power spectrum are not competitive with the CMB for phase transitions occurring after recombination. In the next section, we illustrate this dominance in the concrete case of a first-order phase transition.

\section{First Order Phase transitions after recombination}
\begin{figure*}[t]
    \centering
    \includegraphics[width=0.49\linewidth]{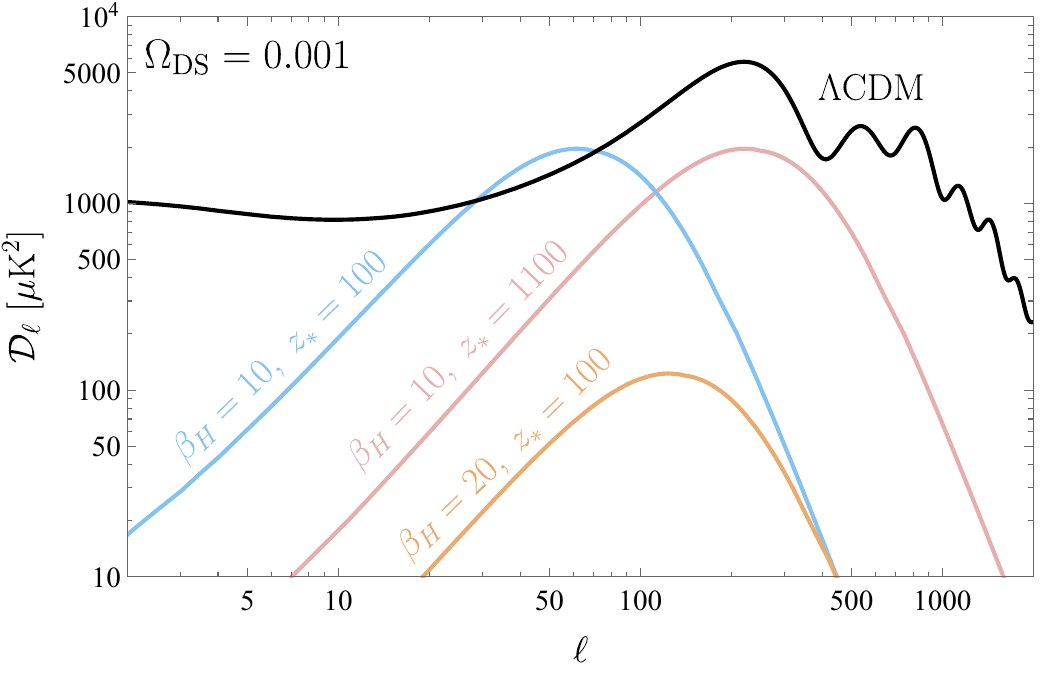}
     \includegraphics[width=0.49\linewidth]{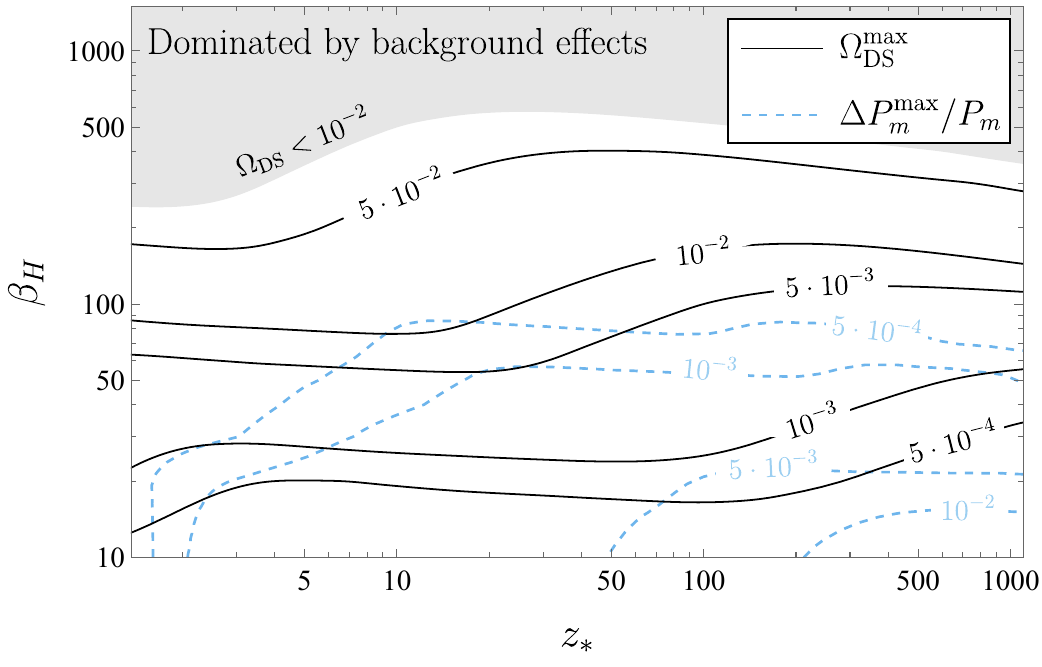}
    \caption{{\bf Left:} Different signal shapes for phase transitions at different redshift of completion $z_*\lesssim1100$ and different bubble size at percolation $1/\beta$. The ratio of this scale to the Hubble radius at $z_*$, encoded by $\beta_H=\beta/H_*$ sets the strength of the phase transition. {\bf Right:} Allowed fraction of energy density at the injection time, $\Omega_{\rm DS}$ ({\bf black solid lines}). For weak phase transitions (large $\beta_H$) modifications of the background dynamics are expected to provide the dominant constraint ({\bf gray shaded region}). We also show the fractional contribution of the late-time generated matter fluctuations to the matter power spectrum today ({\bf light blue dashed lines}).}
    \label{fig:money}
\end{figure*}
For concreteness, we consider a first-order phase transition occurring after recombination. First-order phase transitions have been extensively studied as a source of tensor gravitational waves in the early Universe, and constitute one of the primary targets for beyond-the-Standard-Model searches with gravitational wave interferometers. As a result, numerical and semi-analytical methods to evaluate the transverse-traceless form factor in Eq.~\eqref{eq:ff} have been developed and cross-validated~\cite{Turner:1992tz,Kosowsky:1992vn,Jinno:2016vai,Gould:2019qek}.

By contrast, much less is known about the scalar and vector form factors, and in particular about the anisotropic stress form factor, which dominates the ISW contribution, as argued in the previous section. Assuming that the dominant contribution to the anisotropic stress arises from bubble collisions—as expected for sufficiently cold dark sectors where plasma contributions are suppressed—we compute the anisotropic stress two-point function semi-analytically in the envelope approximation, generalizing the procedure of Ref.~\cite{Jinno:2016vai} as detailed in App.~\ref{app:anisotropicstress}.

In this setup, we find that the instantaneous injection of anisotropic stress provides a parametrically accurate description of the full signal.\footnote{Another source of signal, first studied in \cite{Elor:2023xbz,Koren:2025ymq,Greene:2026gnw}, is induced by fluctuations in the completion time of the phase transition across different Hubble patches. In this case, the released radiation starts redshifting at slightly different times. This amounts to varying \(a_*\) within a window $\delta a_* = a_* \mathcal H_* \delta\eta,
\qquad \delta\eta\sim \beta_*^{-1},$ in \eqref{eq:background}. This gives
\begin{equation}
    \begin{split}
    \delta\rho_z(\vec{x}) &= 4\mathcal H_* \rho_D \delta\eta(\vec{x}) \theta(a-a_*) ,\\
    \delta p_z(\vec{x}) &= \frac{4}{3}\rho_D a_* \mathcal H_* \delta\eta(\vec{x}) \delta(a-a_*) .
    \end{split}
\end{equation}
These fluctuations fit within the general parametrization of \eqref{eq:disc}. However, since they do not source an instantaneous anisotropic-stress injection, they contribute only to the continuous part of the metric response and are therefore subdominant.} Continuous contributions from all other gravitational modes are suppressed by powers of $1/\beta_H$, so that the ISW response is dominated by the impulsive component. The full expression in Eq.~\eqref{eq:ISWfull} therefore reduces to
\begin{equation}
\mathcal{D}_\ell\approx
\frac{\Omega_{\rm DS}^2}{\beta_H^4}
\frac{9\ell(\ell+1)\bar{T}^2}{\pi^2}
\!\int \!\mathrm{d}x\,
\left[j_\ell'(x)\right]^2
P_{\sigma\sigma}\!\left(\frac{x}{\beta_* \chi_*}\right)\, .
\end{equation}
Since $P_{\sigma\sigma}(x)$ approaches a constant at low $x$ and exhibits a bump around $x\simeq 1$, the integral produces a smeared bump in multipole space around $\ell_* \sim \beta_* \chi_*$, as shown in Fig.~\ref{fig:money}.

To constrain the signal, we compare the $\chi^2$ of the $\mathcal{D}_\ell$ measured by Planck\footnote{Data taken from the \href{http://pla.esac.esa.int/pla/\#home}{Planck Legacy Archive}.} assuming the $\Lambda$CDM fit with that where the PT signal is added. We then estimate the 95\% CL exclusion limit on $\Ods$ by requiring $\Delta\chi^2=\chi^2(\Lambda{\rm CDM+PT})-\chi^2(\Lambda{\rm CDM})\simeq2.71$.
The results are shown in the right panel of Fig.~\ref{fig:money}, where contours of the allowed $\Omega_{\rm DS}$ are presented at fixed $z_*$ and $\beta_H$. Increasing $\beta_H$ suppresses the signal as $\beta_H^{-4}$ and shifts the peak to larger multipoles, while increasing $z_*$ shifts the peak according to $\ell_* \propto (1+z_*)^{1/2}$.

At large $\beta_H$, the transition becomes increasingly weak and approaches the second-order regime. In this limit, its dominant effects arise from modifications to the background evolution, which can be constrained using combined CMB and large-scale structure observations following Refs.~\cite{Archidiacono:2022iuu,Bottaro:2023wkd}. More concretely, we estimated the upper limit on $\Ods$ by comparing the correction to the growth of matter fluctuations due to the DS background \eqref{eq:background} with the corresponding correction in the dark-force scenario of~\cite{Archidiacono:2022iuu,Bottaro:2023wkd}.  

Finally, we show in the right panel of Fig.~\ref{fig:money} the contribution of the late-time sourced fluctuations to the present-day matter power spectrum after imposing CMB constraints. The allowed contributions are small across most of the parameter space, with percent-level deviations only at high redshift for sufficiently strong phase transitions.

\section{Outlook}

We developed a general formalism to describe sudden fluctuation injections in the late Universe. For injections after recombination, the phenomenology simplifies, effectively reducing to a search for a bump in the CMB temperature power spectrum induced via the ISW effect by the instantaneous injection of anisotropic stress. As an application, we constrained the parameter space of first-order phase transitions completely sequestered from $\Lambda$CDM during matter domination.

A natural extension of this work is to study phase transitions during dark energy domination, which are expected to display a phenomenology similar to that discussed here, with potentially interesting implications for models of dynamical dark energy. Fluctuation injections during radiation domination offer a complementary regime, where the relative sensitivity of CMB and matter power spectrum observables is less sharply delineated. In this context, CMB polarization, together with its cross-correlation with temperature anisotropies, could help disentangle scalar, vector, and tensor contributions, allowing for a more refined characterization of the underlying dynamics. This effect is generally suppressed during matter domination due to the absence of an efficient photon polarizer, with the notable exception of the reionization epoch. 

These directions would generalize the preliminary studies of Refs.~\cite{Elor:2023xbz,Koren:2025ymq,Greene:2026gnw}, incorporating important subhorizon effects and motivating further theoretical work to compute these sources. A numerical implementation in Boltzmann codes for the CMB and the matter power spectrum~\cite{Blas:2011rf} will also be required to carry out a systematic exploration of this framework.

Concerning smoking-gun signals of these scenarios, the non-Gaussian nature of the injected source will be directly reflected in CMB non-Gaussianities and in the galaxy bispectrum. A computation of the three-point correlators necessary to reliably predict these signals for a first-order phase transition is left for future work. 

\section{Acknowledgments}

We thank Mustafa Amin, Kfir Blum, Mehrdad Mirbabayi, Marko Simonovic and Eleonora Vanzan for many enjoyable discussions. This project was sparked by exchanges during the workshop “New Physics from Galaxy Clustering” at the Galileo Galilei Institute. A large part of this work was carried out at the Galileo Galilei Institute, whose welcoming atmosphere and generous hospitality we gratefully acknowledge. MG is supported by the Israel Science Foundation under Grant No. 1424/23 and by the NSF-BSF grant 2023711.

\bibliography{biblio.bib}

\begin{thebibliography}{31}%
\makeatletter
\providecommand \@ifxundefined [1]{%
 \@ifx{#1\undefined}
}%
\providecommand \@ifnum [1]{%
 \ifnum #1\expandafter \@firstoftwo
 \else \expandafter \@secondoftwo
 \fi
}%
\providecommand \@ifx [1]{%
 \ifx #1\expandafter \@firstoftwo
 \else \expandafter \@secondoftwo
 \fi
}%
\providecommand \natexlab [1]{#1}%
\providecommand \enquote  [1]{``#1''}%
\providecommand \bibnamefont  [1]{#1}%
\providecommand \bibfnamefont [1]{#1}%
\providecommand \citenamefont [1]{#1}%
\providecommand \href@noop [0]{\@secondoftwo}%
\providecommand \href [0]{\begingroup \@sanitize@url \@href}%
\providecommand \@href[1]{\@@startlink{#1}\@@href}%
\providecommand \@@href[1]{\endgroup#1\@@endlink}%
\providecommand \@sanitize@url [0]{\catcode `\\12\catcode `\$12\catcode
  `\&12\catcode `\#12\catcode `\^12\catcode `\_12\catcode `\%12\relax}%
\providecommand \@@startlink[1]{}%
\providecommand \@@endlink[0]{}%
\providecommand \url  [0]{\begingroup\@sanitize@url \@url }%
\providecommand \@url [1]{\endgroup\@href {#1}{\urlprefix }}%
\providecommand \urlprefix  [0]{URL }%
\providecommand \Eprint [0]{\href }%
\providecommand \doibase [0]{http://dx.doi.org/}%
\providecommand \selectlanguage [0]{\@gobble}%
\providecommand \bibinfo  [0]{\@secondoftwo}%
\providecommand \bibfield  [0]{\@secondoftwo}%
\providecommand \translation [1]{[#1]}%
\providecommand \BibitemOpen [0]{}%
\providecommand \bibitemStop [0]{}%
\providecommand \bibitemNoStop [0]{.\EOS\space}%
\providecommand \EOS [0]{\spacefactor3000\relax}%
\providecommand \BibitemShut  [1]{\csname bibitem#1\endcsname}%
\let\auto@bib@innerbib\@empty
\bibitem [{\citenamefont {Kibble}(1980)}]{Kibble1980}%
  \BibitemOpen
  \bibfield  {author} {\bibinfo {author} {\bibfnamefont {T.~W.~B.}\
  \bibnamefont {Kibble}},\ }\href@noop {} {\bibfield  {journal} {\bibinfo
  {journal} {Physics Reports}\ }\textbf {\bibinfo {volume} {67}},\ \bibinfo
  {pages} {183} (\bibinfo {year} {1980})}\BibitemShut {NoStop}%
\bibitem [{\citenamefont {Witten}(1984)}]{Witten1984}%
  \BibitemOpen
  \bibfield  {author} {\bibinfo {author} {\bibfnamefont {E.}~\bibnamefont
  {Witten}},\ }\href@noop {} {\bibfield  {journal} {\bibinfo  {journal}
  {Physical Review D}\ }\textbf {\bibinfo {volume} {30}},\ \bibinfo {pages}
  {272} (\bibinfo {year} {1984})}\BibitemShut {NoStop}%
\bibitem [{\citenamefont {Guth}\ and\ \citenamefont {Tye}(1980)}]{GuthTye1980}%
  \BibitemOpen
  \bibfield  {author} {\bibinfo {author} {\bibfnamefont {A.~H.}\ \bibnamefont
  {Guth}}\ and\ \bibinfo {author} {\bibfnamefont {S.-H.~H.}\ \bibnamefont
  {Tye}},\ }\href@noop {} {\bibfield  {journal} {\bibinfo  {journal} {Physical
  Review Letters}\ }\textbf {\bibinfo {volume} {44}},\ \bibinfo {pages} {631}
  (\bibinfo {year} {1980})}\BibitemShut {NoStop}%
\bibitem [{\citenamefont {Kibble}(1976)}]{Kibble1976}%
  \BibitemOpen
  \bibfield  {author} {\bibinfo {author} {\bibfnamefont {T.~W.~B.}\
  \bibnamefont {Kibble}},\ }\href@noop {} {\bibfield  {journal} {\bibinfo
  {journal} {Journal of Physics A}\ }\textbf {\bibinfo {volume} {9}},\ \bibinfo
  {pages} {1387} (\bibinfo {year} {1976})}\BibitemShut {NoStop}%
\bibitem [{\citenamefont {Zurek}(1985)}]{Zurek1985}%
  \BibitemOpen
  \bibfield  {author} {\bibinfo {author} {\bibfnamefont {W.~H.}\ \bibnamefont
  {Zurek}},\ }\href@noop {} {\bibfield  {journal} {\bibinfo  {journal}
  {Nature}\ }\textbf {\bibinfo {volume} {317}},\ \bibinfo {pages} {505}
  (\bibinfo {year} {1985})}\BibitemShut {NoStop}%
\bibitem [{\citenamefont {Vilenkin}\ and\ \citenamefont
  {Shellard}(1994)}]{VilenkinShellard1994}%
  \BibitemOpen
  \bibfield  {author} {\bibinfo {author} {\bibfnamefont {A.}~\bibnamefont
  {Vilenkin}}\ and\ \bibinfo {author} {\bibfnamefont {E.~P.~S.}\ \bibnamefont
  {Shellard}},\ }\href@noop {} {\emph {\bibinfo {title} {Cosmic Strings and
  Other Topological Defects}}}\ (\bibinfo  {publisher} {Cambridge University
  Press},\ \bibinfo {year} {1994})\BibitemShut {NoStop}%
\bibitem [{\citenamefont {Hindmarsh}\ and\ \citenamefont
  {Kibble}(1995)}]{HindmarshKibble1995}%
  \BibitemOpen
  \bibfield  {author} {\bibinfo {author} {\bibfnamefont {M.}~\bibnamefont
  {Hindmarsh}}\ and\ \bibinfo {author} {\bibfnamefont {T.~W.~B.}\ \bibnamefont
  {Kibble}},\ }\href@noop {} {\bibfield  {journal} {\bibinfo  {journal}
  {Reports on Progress in Physics}\ }\textbf {\bibinfo {volume} {58}},\
  \bibinfo {pages} {477} (\bibinfo {year} {1995})}\BibitemShut {NoStop}%
\bibitem [{\citenamefont {Felder}\ \emph
  {et~al.}(2001{\natexlab{a}})\citenamefont {Felder}, \citenamefont {Kofman},\
  and\ \citenamefont {Linde}}]{FelderKofmanLinde2001PRD}%
  \BibitemOpen
  \bibfield  {author} {\bibinfo {author} {\bibfnamefont {G.}~\bibnamefont
  {Felder}}, \bibinfo {author} {\bibfnamefont {L.}~\bibnamefont {Kofman}}, \
  and\ \bibinfo {author} {\bibfnamefont {A.}~\bibnamefont {Linde}},\
  }\href@noop {} {\bibfield  {journal} {\bibinfo  {journal} {Physical Review
  D}\ }\textbf {\bibinfo {volume} {64}},\ \bibinfo {pages} {123517} (\bibinfo
  {year} {2001}{\natexlab{a}})}\BibitemShut {NoStop}%
\bibitem [{\citenamefont {Felder}\ \emph
  {et~al.}(2001{\natexlab{b}})\citenamefont {Felder}, \citenamefont {Kofman},
  \citenamefont {Linde},\ and\ \citenamefont {Tkachev}}]{FelderEtAl2001PRL}%
  \BibitemOpen
  \bibfield  {author} {\bibinfo {author} {\bibfnamefont {G.}~\bibnamefont
  {Felder}}, \bibinfo {author} {\bibfnamefont {L.}~\bibnamefont {Kofman}},
  \bibinfo {author} {\bibfnamefont {A.}~\bibnamefont {Linde}}, \ and\ \bibinfo
  {author} {\bibfnamefont {I.}~\bibnamefont {Tkachev}},\ }\href@noop {}
  {\bibfield  {journal} {\bibinfo  {journal} {Physical Review Letters}\
  }\textbf {\bibinfo {volume} {87}},\ \bibinfo {pages} {011601} (\bibinfo
  {year} {2001}{\natexlab{b}})}\BibitemShut {NoStop}%
\bibitem [{\citenamefont {Parker}(1968)}]{Parker1968}%
  \BibitemOpen
  \bibfield  {author} {\bibinfo {author} {\bibfnamefont {L.}~\bibnamefont
  {Parker}},\ }\href@noop {} {\bibfield  {journal} {\bibinfo  {journal}
  {Physical Review Letters}\ }\textbf {\bibinfo {volume} {21}},\ \bibinfo
  {pages} {562} (\bibinfo {year} {1968})}\BibitemShut {NoStop}%
\bibitem [{\citenamefont {Kofman}\ \emph {et~al.}(1994)\citenamefont {Kofman},
  \citenamefont {Linde},\ and\ \citenamefont
  {Starobinsky}}]{KofmanLindeStarobinsky1994}%
  \BibitemOpen
  \bibfield  {author} {\bibinfo {author} {\bibfnamefont {L.}~\bibnamefont
  {Kofman}}, \bibinfo {author} {\bibfnamefont {A.}~\bibnamefont {Linde}}, \
  and\ \bibinfo {author} {\bibfnamefont {A.}~\bibnamefont {Starobinsky}},\
  }\href@noop {} {\bibfield  {journal} {\bibinfo  {journal} {Physical Review
  Letters}\ }\textbf {\bibinfo {volume} {73}},\ \bibinfo {pages} {3195}
  (\bibinfo {year} {1994})}\BibitemShut {NoStop}%
\bibitem [{\citenamefont {Kofman}\ \emph {et~al.}(1997)\citenamefont {Kofman},
  \citenamefont {Linde},\ and\ \citenamefont
  {Starobinsky}}]{KofmanLindeStarobinsky1997}%
  \BibitemOpen
  \bibfield  {author} {\bibinfo {author} {\bibfnamefont {L.}~\bibnamefont
  {Kofman}}, \bibinfo {author} {\bibfnamefont {A.}~\bibnamefont {Linde}}, \
  and\ \bibinfo {author} {\bibfnamefont {A.}~\bibnamefont {Starobinsky}},\
  }\href@noop {} {\bibfield  {journal} {\bibinfo  {journal} {Physical Review
  D}\ }\textbf {\bibinfo {volume} {56}},\ \bibinfo {pages} {3258} (\bibinfo
  {year} {1997})}\BibitemShut {NoStop}%
\bibitem [{\citenamefont {Mazumdar}\ and\ \citenamefont
  {White}(2019)}]{MazumdarWhite2019}%
  \BibitemOpen
  \bibfield  {author} {\bibinfo {author} {\bibfnamefont {A.}~\bibnamefont
  {Mazumdar}}\ and\ \bibinfo {author} {\bibfnamefont {G.}~\bibnamefont
  {White}},\ }\href@noop {} {\bibfield  {journal} {\bibinfo  {journal} {Reports
  on Progress in Physics}\ }\textbf {\bibinfo {volume} {82}},\ \bibinfo {pages}
  {076901} (\bibinfo {year} {2019})}\BibitemShut {NoStop}%
\bibitem [{\citenamefont {Amin}\ \emph {et~al.}(2014)\citenamefont {Amin},
  \citenamefont {Hertzberg}, \citenamefont {Kaiser},\ and\ \citenamefont
  {Karouby}}]{AminEtAl2014}%
  \BibitemOpen
  \bibfield  {author} {\bibinfo {author} {\bibfnamefont {M.~A.}\ \bibnamefont
  {Amin}}, \bibinfo {author} {\bibfnamefont {M.~P.}\ \bibnamefont {Hertzberg}},
  \bibinfo {author} {\bibfnamefont {D.~I.}\ \bibnamefont {Kaiser}}, \ and\
  \bibinfo {author} {\bibfnamefont {J.}~\bibnamefont {Karouby}},\ }\href@noop
  {} {\bibfield  {journal} {\bibinfo  {journal} {International Journal of
  Modern Physics D}\ }\textbf {\bibinfo {volume} {24}},\ \bibinfo {pages}
  {1530003} (\bibinfo {year} {2014})}\BibitemShut {NoStop}%
\bibitem [{\citenamefont {Creminelli}\ \emph {et~al.}(2009)\citenamefont
  {Creminelli}, \citenamefont {D'Amico}, \citenamefont {Norena},\ and\
  \citenamefont {Vernizzi}}]{Creminelli:2008wc}%
  \BibitemOpen
  \bibfield  {author} {\bibinfo {author} {\bibfnamefont {P.}~\bibnamefont
  {Creminelli}}, \bibinfo {author} {\bibfnamefont {G.}~\bibnamefont {D'Amico}},
  \bibinfo {author} {\bibfnamefont {J.}~\bibnamefont {Norena}}, \ and\ \bibinfo
  {author} {\bibfnamefont {F.}~\bibnamefont {Vernizzi}},\ }\href {\doibase
  10.1088/1475-7516/2009/02/018} {\bibfield  {journal} {\bibinfo  {journal}
  {JCAP}\ }\textbf {\bibinfo {volume} {02}},\ \bibinfo {pages} {018} (\bibinfo
  {year} {2009})},\ \Eprint {http://arxiv.org/abs/0811.0827} {arXiv:0811.0827
  [astro-ph]} \BibitemShut {NoStop}%
\bibitem [{\citenamefont {Jinno}\ and\ \citenamefont
  {Takimoto}(2017)}]{Jinno:2016vai}%
  \BibitemOpen
  \bibfield  {author} {\bibinfo {author} {\bibfnamefont {R.}~\bibnamefont
  {Jinno}}\ and\ \bibinfo {author} {\bibfnamefont {M.}~\bibnamefont
  {Takimoto}},\ }\href {\doibase 10.1103/PhysRevD.95.024009} {\bibfield
  {journal} {\bibinfo  {journal} {Phys. Rev. D}\ }\textbf {\bibinfo {volume}
  {95}},\ \bibinfo {pages} {024009} (\bibinfo {year} {2017})},\ \Eprint
  {http://arxiv.org/abs/1605.01403} {arXiv:1605.01403 [astro-ph.CO]}
  \BibitemShut {NoStop}%
\bibitem [{\citenamefont {Turner}\ \emph {et~al.}(1992)\citenamefont {Turner},
  \citenamefont {Weinberg},\ and\ \citenamefont {Widrow}}]{Turner:1992tz}%
  \BibitemOpen
  \bibfield  {author} {\bibinfo {author} {\bibfnamefont {M.~S.}\ \bibnamefont
  {Turner}}, \bibinfo {author} {\bibfnamefont {E.~J.}\ \bibnamefont
  {Weinberg}}, \ and\ \bibinfo {author} {\bibfnamefont {L.~M.}\ \bibnamefont
  {Widrow}},\ }\href {\doibase 10.1103/PhysRevD.46.2384} {\bibfield  {journal}
  {\bibinfo  {journal} {Phys. Rev. D}\ }\textbf {\bibinfo {volume} {46}},\
  \bibinfo {pages} {2384} (\bibinfo {year} {1992})}\BibitemShut {NoStop}%
\bibitem [{\citenamefont {Caprini}\ \emph {et~al.}(2009)\citenamefont
  {Caprini}, \citenamefont {Durrer}, \citenamefont {Konstandin},\ and\
  \citenamefont {Servant}}]{Caprini:2009fx}%
  \BibitemOpen
  \bibfield  {author} {\bibinfo {author} {\bibfnamefont {C.}~\bibnamefont
  {Caprini}}, \bibinfo {author} {\bibfnamefont {R.}~\bibnamefont {Durrer}},
  \bibinfo {author} {\bibfnamefont {T.}~\bibnamefont {Konstandin}}, \ and\
  \bibinfo {author} {\bibfnamefont {G.}~\bibnamefont {Servant}},\ }\href
  {\doibase 10.1103/PhysRevD.79.083519} {\bibfield  {journal} {\bibinfo
  {journal} {Phys. Rev. D}\ }\textbf {\bibinfo {volume} {79}},\ \bibinfo
  {pages} {083519} (\bibinfo {year} {2009})},\ \Eprint
  {http://arxiv.org/abs/0901.1661} {arXiv:0901.1661 [astro-ph.CO]} \BibitemShut
  {NoStop}%
\bibitem [{\citenamefont {Ma}\ and\ \citenamefont
  {Bertschinger}(1995)}]{Ma:1995ey}%
  \BibitemOpen
  \bibfield  {author} {\bibinfo {author} {\bibfnamefont {C.-P.}\ \bibnamefont
  {Ma}}\ and\ \bibinfo {author} {\bibfnamefont {E.}~\bibnamefont
  {Bertschinger}},\ }\href {\doibase 10.1086/176550} {\bibfield  {journal}
  {\bibinfo  {journal} {Astrophys. J.}\ }\textbf {\bibinfo {volume} {455}},\
  \bibinfo {pages} {7} (\bibinfo {year} {1995})},\ \Eprint
  {http://arxiv.org/abs/astro-ph/9506072} {arXiv:astro-ph/9506072} \BibitemShut
  {NoStop}%
\bibitem [{\citenamefont {Gorbunov}\ and\ \citenamefont
  {Rubakov}(2011)}]{Gorbunov:2011zzc}%
  \BibitemOpen
  \bibfield  {author} {\bibinfo {author} {\bibfnamefont {D.~S.}\ \bibnamefont
  {Gorbunov}}\ and\ \bibinfo {author} {\bibfnamefont {V.~A.}\ \bibnamefont
  {Rubakov}},\ }\href {\doibase 10.1142/7873} {\emph {\bibinfo {title}
  {{Introduction to the theory of the early universe: Cosmological
  perturbations and inflationary theory}}}}\ (\bibinfo {year}
  {2011})\BibitemShut {NoStop}%
\bibitem [{\citenamefont {LoVerde}\ and\ \citenamefont
  {Afshordi}(2008)}]{LoVerde:2008re}%
  \BibitemOpen
  \bibfield  {author} {\bibinfo {author} {\bibfnamefont {M.}~\bibnamefont
  {LoVerde}}\ and\ \bibinfo {author} {\bibfnamefont {N.}~\bibnamefont
  {Afshordi}},\ }\href {\doibase 10.1103/PhysRevD.78.123506} {\bibfield
  {journal} {\bibinfo  {journal} {Phys. Rev. D}\ }\textbf {\bibinfo {volume}
  {78}},\ \bibinfo {pages} {123506} (\bibinfo {year} {2008})},\ \Eprint
  {http://arxiv.org/abs/0809.5112} {arXiv:0809.5112 [astro-ph]} \BibitemShut
  {NoStop}%
\bibitem [{\citenamefont {Kosowsky}\ and\ \citenamefont
  {Turner}(1993)}]{Kosowsky:1992vn}%
  \BibitemOpen
  \bibfield  {author} {\bibinfo {author} {\bibfnamefont {A.}~\bibnamefont
  {Kosowsky}}\ and\ \bibinfo {author} {\bibfnamefont {M.~S.}\ \bibnamefont
  {Turner}},\ }\href {\doibase 10.1103/PhysRevD.47.4372} {\bibfield  {journal}
  {\bibinfo  {journal} {Phys. Rev. D}\ }\textbf {\bibinfo {volume} {47}},\
  \bibinfo {pages} {4372} (\bibinfo {year} {1993})},\ \Eprint
  {http://arxiv.org/abs/astro-ph/9211004} {arXiv:astro-ph/9211004} \BibitemShut
  {NoStop}%
\bibitem [{\citenamefont {Gould}\ \emph {et~al.}(2019)\citenamefont {Gould},
  \citenamefont {Kozaczuk}, \citenamefont {Niemi}, \citenamefont
  {Ramsey-Musolf}, \citenamefont {Tenkanen},\ and\ \citenamefont
  {Weir}}]{Gould:2019qek}%
  \BibitemOpen
  \bibfield  {author} {\bibinfo {author} {\bibfnamefont {O.}~\bibnamefont
  {Gould}}, \bibinfo {author} {\bibfnamefont {J.}~\bibnamefont {Kozaczuk}},
  \bibinfo {author} {\bibfnamefont {L.}~\bibnamefont {Niemi}}, \bibinfo
  {author} {\bibfnamefont {M.~J.}\ \bibnamefont {Ramsey-Musolf}}, \bibinfo
  {author} {\bibfnamefont {T.~V.~I.}\ \bibnamefont {Tenkanen}}, \ and\ \bibinfo
  {author} {\bibfnamefont {D.~J.}\ \bibnamefont {Weir}},\ }\href {\doibase
  10.1103/PhysRevD.100.115024} {\bibfield  {journal} {\bibinfo  {journal}
  {Phys. Rev. D}\ }\textbf {\bibinfo {volume} {100}},\ \bibinfo {pages}
  {115024} (\bibinfo {year} {2019})},\ \Eprint
  {http://arxiv.org/abs/1903.11604} {arXiv:1903.11604 [hep-ph]} \BibitemShut
  {NoStop}%
\bibitem [{\citenamefont {Elor}\ \emph {et~al.}(2024)\citenamefont {Elor},
  \citenamefont {Jinno}, \citenamefont {Kumar}, \citenamefont {McGehee},\ and\
  \citenamefont {Tsai}}]{Elor:2023xbz}%
  \BibitemOpen
  \bibfield  {author} {\bibinfo {author} {\bibfnamefont {G.}~\bibnamefont
  {Elor}}, \bibinfo {author} {\bibfnamefont {R.}~\bibnamefont {Jinno}},
  \bibinfo {author} {\bibfnamefont {S.}~\bibnamefont {Kumar}}, \bibinfo
  {author} {\bibfnamefont {R.}~\bibnamefont {McGehee}}, \ and\ \bibinfo
  {author} {\bibfnamefont {Y.}~\bibnamefont {Tsai}},\ }\href {\doibase
  10.1103/PhysRevLett.133.211003} {\bibfield  {journal} {\bibinfo  {journal}
  {Phys. Rev. Lett.}\ }\textbf {\bibinfo {volume} {133}},\ \bibinfo {pages}
  {211003} (\bibinfo {year} {2024})},\ \Eprint
  {http://arxiv.org/abs/2311.16222} {arXiv:2311.16222 [hep-ph]} \BibitemShut
  {NoStop}%
\bibitem [{\citenamefont {Koren}\ \emph {et~al.}(2025)\citenamefont {Koren},
  \citenamefont {Tsai},\ and\ \citenamefont {Wang}}]{Koren:2025ymq}%
  \BibitemOpen
  \bibfield  {author} {\bibinfo {author} {\bibfnamefont {S.}~\bibnamefont
  {Koren}}, \bibinfo {author} {\bibfnamefont {Y.}~\bibnamefont {Tsai}}, \ and\
  \bibinfo {author} {\bibfnamefont {R.}~\bibnamefont {Wang}},\ }\href@noop {}
  {\  (\bibinfo {year} {2025})},\ \Eprint {http://arxiv.org/abs/2509.07076}
  {arXiv:2509.07076 [hep-ph]} \BibitemShut {NoStop}%
\bibitem [{\citenamefont {Greene}\ \emph {et~al.}(2026)\citenamefont {Greene},
  \citenamefont {Ho}, \citenamefont {Kumar},\ and\ \citenamefont
  {Tsai}}]{Greene:2026gnw}%
  \BibitemOpen
  \bibfield  {author} {\bibinfo {author} {\bibfnamefont {K.}~\bibnamefont
  {Greene}}, \bibinfo {author} {\bibfnamefont {D.~W.~R.}\ \bibnamefont {Ho}},
  \bibinfo {author} {\bibfnamefont {S.}~\bibnamefont {Kumar}}, \ and\ \bibinfo
  {author} {\bibfnamefont {Y.}~\bibnamefont {Tsai}},\ }\href@noop {} {\
  (\bibinfo {year} {2026})},\ \Eprint {http://arxiv.org/abs/2603.00272}
  {arXiv:2603.00272 [hep-ph]} \BibitemShut {NoStop}%
\bibitem [{\citenamefont {Archidiacono}\ \emph {et~al.}(2022)\citenamefont
  {Archidiacono}, \citenamefont {Castorina}, \citenamefont {Redigolo},\ and\
  \citenamefont {Salvioni}}]{Archidiacono:2022iuu}%
  \BibitemOpen
  \bibfield  {author} {\bibinfo {author} {\bibfnamefont {M.}~\bibnamefont
  {Archidiacono}}, \bibinfo {author} {\bibfnamefont {E.}~\bibnamefont
  {Castorina}}, \bibinfo {author} {\bibfnamefont {D.}~\bibnamefont {Redigolo}},
  \ and\ \bibinfo {author} {\bibfnamefont {E.}~\bibnamefont {Salvioni}},\
  }\href {\doibase 10.1088/1475-7516/2022/10/074} {\bibfield  {journal}
  {\bibinfo  {journal} {JCAP}\ }\textbf {\bibinfo {volume} {10}},\ \bibinfo
  {pages} {074} (\bibinfo {year} {2022})},\ \Eprint
  {http://arxiv.org/abs/2204.08484} {arXiv:2204.08484 [astro-ph.CO]}
  \BibitemShut {NoStop}%
\bibitem [{\citenamefont {Bottaro}\ \emph {et~al.}(2024)\citenamefont
  {Bottaro}, \citenamefont {Castorina}, \citenamefont {Costa}, \citenamefont
  {Redigolo},\ and\ \citenamefont {Salvioni}}]{Bottaro:2023wkd}%
  \BibitemOpen
  \bibfield  {author} {\bibinfo {author} {\bibfnamefont {S.}~\bibnamefont
  {Bottaro}}, \bibinfo {author} {\bibfnamefont {E.}~\bibnamefont {Castorina}},
  \bibinfo {author} {\bibfnamefont {M.}~\bibnamefont {Costa}}, \bibinfo
  {author} {\bibfnamefont {D.}~\bibnamefont {Redigolo}}, \ and\ \bibinfo
  {author} {\bibfnamefont {E.}~\bibnamefont {Salvioni}},\ }\href {\doibase
  10.1103/PhysRevLett.132.201002} {\bibfield  {journal} {\bibinfo  {journal}
  {Phys. Rev. Lett.}\ }\textbf {\bibinfo {volume} {132}},\ \bibinfo {pages}
  {201002} (\bibinfo {year} {2024})},\ \Eprint
  {http://arxiv.org/abs/2309.11496} {arXiv:2309.11496 [astro-ph.CO]}
  \BibitemShut {NoStop}%
\bibitem [{\citenamefont {Blas}\ \emph {et~al.}(2011)\citenamefont {Blas},
  \citenamefont {Lesgourgues},\ and\ \citenamefont {Tram}}]{Blas:2011rf}%
  \BibitemOpen
  \bibfield  {author} {\bibinfo {author} {\bibfnamefont {D.}~\bibnamefont
  {Blas}}, \bibinfo {author} {\bibfnamefont {J.}~\bibnamefont {Lesgourgues}}, \
  and\ \bibinfo {author} {\bibfnamefont {T.}~\bibnamefont {Tram}},\ }\href
  {\doibase 10.1088/1475-7516/2011/07/034} {\bibfield  {journal} {\bibinfo
  {journal} {JCAP}\ }\textbf {\bibinfo {volume} {07}},\ \bibinfo {pages} {034}
  (\bibinfo {year} {2011})},\ \Eprint {http://arxiv.org/abs/1104.2933}
  {arXiv:1104.2933 [astro-ph.CO]} \BibitemShut {NoStop}%
\bibitem [{\citenamefont {Bottaro}\ \emph {et~al.}(pear)\citenamefont
  {Bottaro}, \citenamefont {Cipressi}, \citenamefont {Hatzofe},\ and\
  \citenamefont {Opferkuch}}]{toappear}%
  \BibitemOpen
  \bibfield  {author} {\bibinfo {author} {\bibfnamefont {S.}~\bibnamefont
  {Bottaro}}, \bibinfo {author} {\bibfnamefont {M.}~\bibnamefont {Cipressi}},
  \bibinfo {author} {\bibfnamefont {B.}~\bibnamefont {Hatzofe}}, \ and\
  \bibinfo {author} {\bibfnamefont {T.}~\bibnamefont {Opferkuch}},\ }\href@noop
  {} {\  (\bibinfo {year} {to appear})}\BibitemShut {NoStop}%
\bibitem [{\citenamefont {Amin}\ \emph {et~al.}(2026)\citenamefont {Amin},
  \citenamefont {Delos},\ and\ \citenamefont {Mirbabayi}}]{Amin:2025dtd}%
  \BibitemOpen
  \bibfield  {author} {\bibinfo {author} {\bibfnamefont {M.~A.}\ \bibnamefont
  {Amin}}, \bibinfo {author} {\bibfnamefont {M.~S.}\ \bibnamefont {Delos}}, \
  and\ \bibinfo {author} {\bibfnamefont {M.}~\bibnamefont {Mirbabayi}},\ }\href
  {\doibase 10.1088/1475-7516/2026/01/016} {\bibfield  {journal} {\bibinfo
  {journal} {JCAP}\ }\textbf {\bibinfo {volume} {01}},\ \bibinfo {pages} {016}
  (\bibinfo {year} {2026})},\ \Eprint {http://arxiv.org/abs/2503.20881}
  {arXiv:2503.20881 [astro-ph.CO]} \BibitemShut {NoStop}%
\end{thebibliography}%

\clearpage
\newpage
\appendix
\onecolumngrid

\section{Transfer functions}\label{sec:fulltransfer}

\begin{figure}[htp!]
    \centering
    \includegraphics[width=0.45\linewidth]{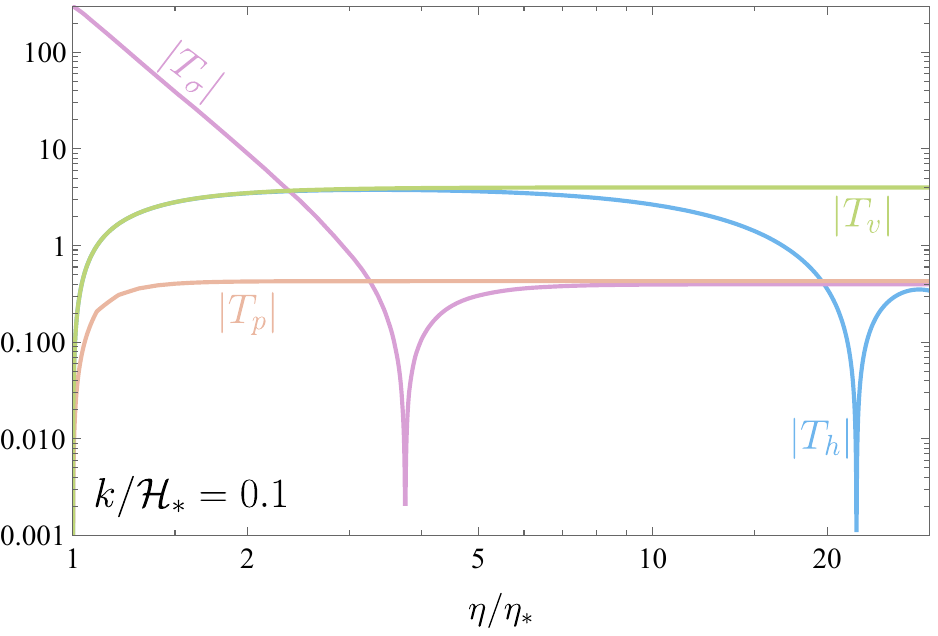}\qquad\includegraphics[width=0.45\linewidth]{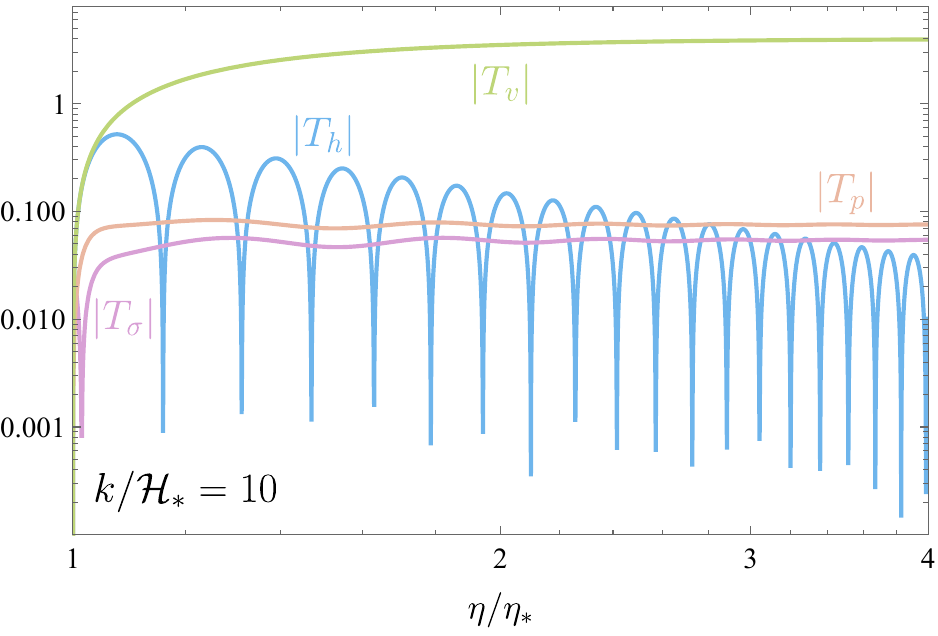}
    \caption{Pressure ({\bf orange}), anisotropic stress ({\bf pink}), vector ({\bf green}) and tensor ({\bf blue}) transfer functions as defined in Eq.~\eqref{eq:psi} as a function of time, normalized to $\eta_*$, for a  superhorizon mode with $k/\mathcal{H}_*<1$ ({\bf left}) and subhorizon mode with $k/\mathcal{H}_*>1$ ({\bf right}) .}
    \label{fig:transfer}
\end{figure}

Here, we go through the derivation of the transfer functions that relate the sources to the metric fluctuations whose definition is repeated here for simplicity 
\begin{align}
&\Psi(\eta,k)=\frac{\Omega_{\rm{DS}}}{\beta_H}\left[ T_\sigma(\eta, k) \hat{\sigma}(\vec{k})+T_p(\eta,k) \hat{p}(\vec{k})\right] \, ,  \label{eq:psi} \\
&W_{i}(\eta,k)=\frac{\Omega_{\rm{DS}}}{\beta_H} T_v(\eta) \hat{\pi}_i^T(\vec{k})\, ,\\
&h_{ij}(\eta,k)= \frac{\Omega_{\rm{DS}}}{\beta_H}T_h(\eta,k) \hat{\pi}_{ij}^{TT}(\vec{k})\, .
\end{align}
Where possible we will provide the analytical expressions for the transfer functions whose behavior as a function of time for super-horizon and sub-horizon modes is plotted in Fig.~\eqref{fig:transfer}. 

\paragraph{Tensor} modes are governed by a single Einstein's equation, which is a wave equation sourced by the transverse traceless part of the spatial stress tensor
\begin{equation}
h_{ij}''+2\ch h_{ij}'+k^2h_{ij}=-\frac{6\ch_*\Ods}{\beta_H}\hat{\pi}^{TT}_{ij}(\vec{k})\delta(\eta-\eta_*)\,.
\end{equation}
The transfer function can then be directly inferred from the homogeneous solutions of the wave equation which is fixed by continuity and the jump condition of the first derivative. All in all we get for $\eta>\eta_*$ 
\begin{equation}
T_h(\eta,k)=12\frac{\eta_*^2}{\eta k}(j_1(k\eta)y_1(k\eta_*)- j_1(k\eta_*)y_1(k\eta))\, ,
\end{equation}
where $j_1(x)=\sin x/x^2-\cos x/x$ and  $y_1(x)=-\cos x/x^2-\sin x/x$ are modified Bessel's functions. Their behavior at large $x$ sets the behavior of the transfer function which  
for $k\eta\gg1$ is a free oscillating wave with frequency $k$ and decaying amplitude 
\begin{equation}
T_h(\eta,k)\approx -12\frac{\eta_*}{\eta}\frac{\sin(k(\eta-\eta_*))}{k \eta }\ .   
\end{equation}
\paragraph{Vector} modes are also evolving following a single Einstein's equation. The assumption that the DS is sequestered from the SM makes it impossible to support vector fluctuation in cosmlogy that, if injected, will have to decay in time following the evolution equation   
\begin{equation}
    \partial_\eta^2W_i^T+2\ch \partial_\eta W_i^T=-\frac{6\ch_*\Ods}{\beta_H}\hat{\pi}^T_{i}(\vec{k})\delta(\eta-\eta_*)\, ,
\end{equation}
where the left hand side can be rewritten in matter domination as $\partial_\eta (\eta^4 \partial_\eta W_i^T)$ so that the transfer function is easily obtained to be for $\eta>\eta_*$
\begin{equation}
T_v(\eta)=4\left[\left(\frac{\eta_*}{\eta}\right)^{3}-1\right]\, .
\end{equation}
This shows how istantaneously injected vector modes quickly decay $\propto\eta^{-3}$. As a consistency check, we show that the dynamics of the vector component $\hat{v}^T_i$ in $T^0_i$ is fixed in terms of the one of $W_i^T$ by the continuity equation and the $0i$ Einstein's equation: 
\begin{equation}
\left\{
    \begin{split}
    & \partial_\eta v_i^T+4\ch v_i^T-k^2\pi^T_i=0\\
        & k^2\partial_\eta W_i^T=-\frac{6\Ods\ch_*^2}{\beta_H}\left(\frac{a}{a_*}\right)^2\hat{v}_i^T\theta(\eta-\eta_*)
    \end{split}\, .
    \right.
\end{equation}
In the context of the phase transitions, this equation describes how $\pi^T_i$ generated by the bubbles produces the vector perturbation $v_i^T$ that will survive in the form of radiation after the PT as 
\begin{equation}
\hat{v}^T_i(\vec{k},\eta)=\left(\frac{\eta_*}{\eta}\right)^8\frac{\eta_*k^2}{2}\hat{\pi}^T_{i}(\hat{k})\ .
\end{equation}

\paragraph{Scalar} Newtonian potential $\Phi$ and the spatial curvature $\Psi$ evolve according to the following Einstein's equations
\begin{equation}\label{eq:einst_scalar}
\left\{
    \begin{split}
        & k^2\Psi+3\ch\Psi'-3\ch^2\Phi=\frac{3}{2}\ch^2\left(\frac{\Ods}{\beta_H}\hat{\rho}\theta(\eta-\eta_*)+\delta_m\right)\\
        & k^2(\Phi+\Psi)=\frac{3\ch_*\Ods}{\beta_H}\hat{\sigma}\delta(\eta-\eta_*)
    \end{split}\, ,
    \right.
\end{equation}
which imply that the structure of the scalar fluctuations is
\begin{equation}
    \begin{split}\label{eq:scalar_decomposition}
        & \Phi(\vec{k},\eta)=\frac{3\Ods}{\ch_*\beta_H}\frac{\ch_*^2}{k^2}\hat{\sigma}(\vec{k})\delta(\eta-\eta_*)-\frac{3\Ods}{\beta_H}\frac{\ch_*^2}{k^2}\hat{\sigma}(\vec{k})\theta(\eta-\eta_*)+c_\Phi^{\rm cont}(\vec{k},\eta)\, ,\\
        & \Psi(\vec{k},\eta)=\frac{3\Ods}{\beta_H}\frac{\ch_*^2}{k^2}\hat{\sigma}(\vec{k})\theta(\eta-\eta_*)+c_\Psi^{\rm cont}(\vec{k},\eta)\ ,
    \end{split}
\end{equation}
with $c_\Psi^{\rm cont}(\vec{k},\eta_*)=c_\Phi^{\rm cont}(\vec{k},\eta_*)=0$. The induced matter fluctuations are described by the set of continuity and Euler's equations
\begin{equation}
    \left\{
\begin{split}
    &\delta_m'+\theta_m+3\Psi'=0\\
    &\theta_m'+\ch\theta_m-k^2\Phi=0
\end{split}\, ,
    \right.
\end{equation}

with boundary conditions

\begin{equation}
    \delta_m(\vec{k})=-\frac{9\Ods}{\beta_H}\frac{\ch_*^2}{k^2}\hat{\sigma}(\vec{k}),\quad \theta_m(\vec{k})=\frac{3\Ods\ch_*}{\beta_H}\hat{\sigma}(\vec{k})\ .
\end{equation}

The equations for the energy density fluctuations in the dark sector are

\begin{equation}\label{eq:fluctuations_rad}
\left\{
    \begin{split}
        &\hat{\rho}'+3\ch(\hatr+\hatp)+\hat{\theta}+3\beta_H(1+w_>)\left(\frac{a_*}{a}\right)^{3(1+w_>)}\Psi'=0\\
        & \hat{\theta}'+4\ch \hat{\theta}-k^2(\hatp+\hats)-\beta_H(1+w_>)\left(\frac{a_*}{a}\right)^{3(1+w_>)}k^2\Phi=0
    \end{split}\,.
    \right.
\end{equation}
where we have already factored out $\rD$ and used the fact that the dark sector is completely secluded from the $\Lambda$CDM sector, including the subdominant radiation component. Notice that the potentials in \eqref{eq:fluctuations_rad} are those generated by the adiabatic initial conditions after inflation, since $\rD$ terms correspond to $\mathcal{O}(\rD^2)$ corrections. The $\Lambda$CDM potentials are continuous, therefore the initial conditions for $\hatr$ and $\hat{\theta}$ are 

\begin{equation}
    \hatr(\vec{k})=-3\hatp(\vec{k}),\quad \hat{\theta}(\vec{k})=\frac{k^2}{\ch_*}\left(\hatp(\vec{k})+\hats(\vec{k})\right)\ .
\end{equation}

Given the set of initial conditions, we can now determine the evolution of the fluctuations for $\eta\geq \eta_*$. Let us assume that the DS is made of radiation, therefore $w_>=1/3$ and $\hatp=\hatr/3$. If we redefine $\hatr=\left(\frac{a_*}{a}\right)^4\hatr_>,\quad \hat{\theta}=\left(\frac{a_*}{a}\right)^4\hat{\theta}_>$ the equations in \eqref{eq:fluctuations_rad} read
\begin{equation}
    \left\{
\begin{split}
   & \hatr_>'+\hat{\theta}_>+4\beta_H\Psi'_{\rm \Lambda CDM}=0\\
   & \hat{\theta}_>'-\frac{k^2}{3}\hatr_>+\frac{4}{3}k^2\beta_H\Psi_{\rm \Lambda CDM}=0
\end{split}\, ,
    \right.
\end{equation}
which, combined, give
\begin{equation}
    \hatr_>''+\frac{k^2}{3}\hatr_>+4\beta_H\Psi''_{\rm \Lambda CDM}-\frac{4}{3}k^2\beta_H\Psi_{\rm \Lambda CDM}=0\ .
\end{equation}
The potentials in $\Lambda$CDM in matter domination are constant, so we get the following solution
\begin{equation}
    \hatr_>=-\left(3\hatp+4\beta_H\Psi_{\rm \Lambda CDM}\right)\cos\left(\frac{k}{\sqrt{3}}(\eta-\eta_*)\right)-\frac{\sqrt{3}k\eta_*}{2}(\hatp+\hats)\sin\left(\frac{k}{\sqrt{3}}(\eta-\eta_*)\right)+4\beta_H\Psi_{\rm \Lambda CDM}\ .
\end{equation}

Next, we determine the evolution of matter fluctuations. If we define $A_m=\delta_m+3\Psi$, we get and using the first equation in \eqref{eq:einst_scalar}, we get

\begin{equation}
    \left\{
    \begin{split}
        &A_m''+\ch A_m'-k^2\Psi=0\, ,\\
        &k^2\Psi+3\ch\Psi'+\frac{15}{2}\ch^2\Psi-\frac{3}{2}\ch^2 A_m=  ~\frac{3}{2}\frac{\ch^2\Ods}{\beta_H}\frac{a_*}{a}\hatr_>\ ,
    \end{split}
    \right.
\end{equation}
where now $\hatr_>$ acts as a source. The new boundary conditions given by
\begin{equation}
    \Psi(\eta_*)=\frac{3\Ods}{\beta_H}\frac{\ch_*^2}{k^2}\hats(\vec{k}),\quad A_m(\eta_*)=0,\quad A_m'(\eta_*)=-\frac{3\Ods}{\beta_H}\ch_*\hats(\vec{k})\ .
\end{equation}

\begin{figure}[htp!]
    \centering
    \includegraphics[width=0.45\linewidth]{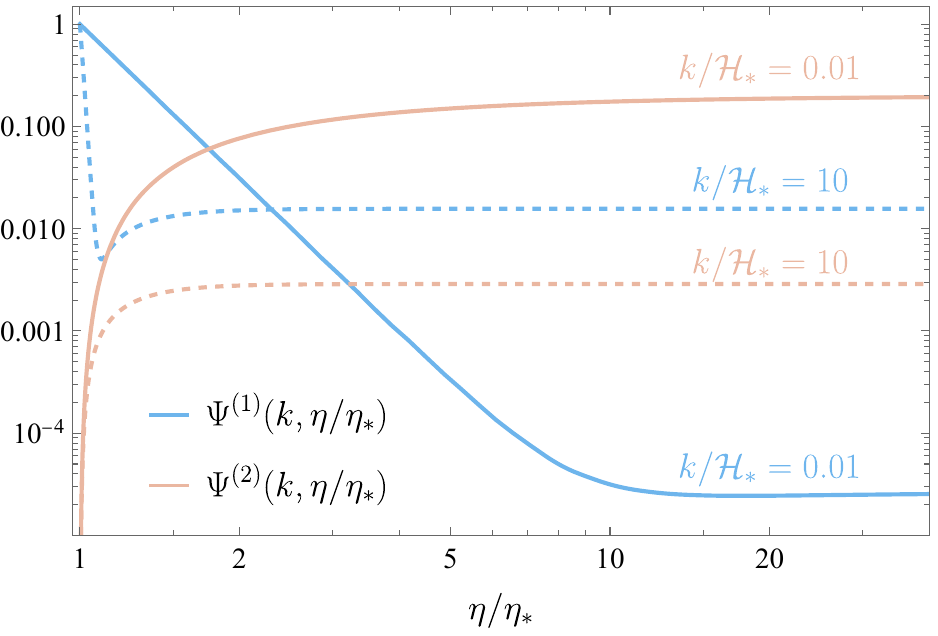}\qquad\includegraphics[width=0.45\linewidth]{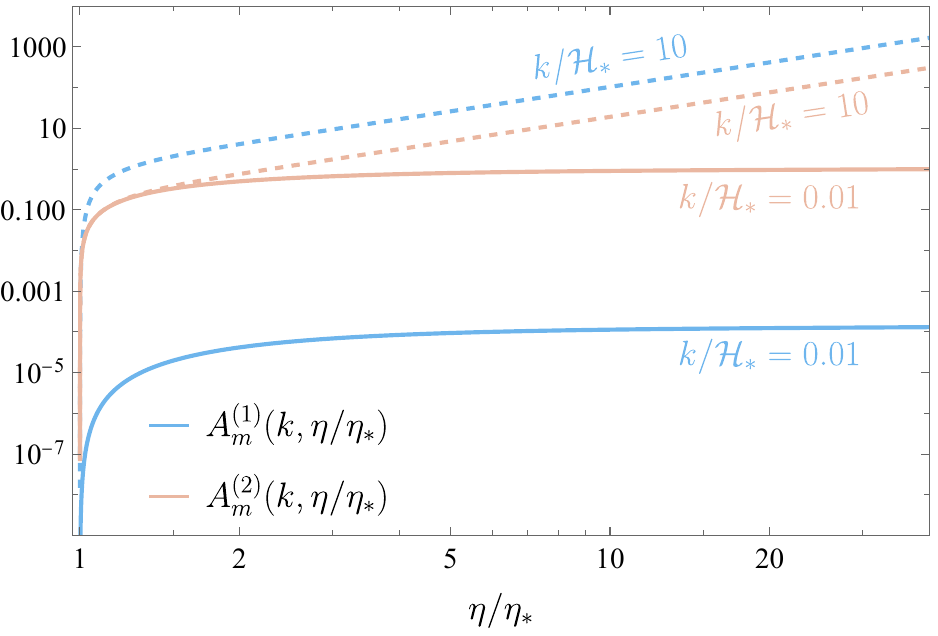}
    \caption{Green's functions $\Psi^{(i)}$ ({\bf left}) and $A_m^{(i)}$ ({\bf right}) obtained by numerically solving \eqref{eq:boltz_einst_y} for a choice of subhorizon e superhorizon mode.}
    \label{fig:green}
\end{figure}

We define two sets of Green's functions in Fourier space that solve

\begin{equation}\label{eq:boltz_einst_y}
\left\{
\begin{split}
    &\frac{1}{\eta}\Psi'+\frac{5}{\eta^2}\Psi-\frac{1}{\eta^2}A_m+\frac{k^2\Psi}{6}=\frac{c^{(1)}}{\eta'}\delta(\eta-\eta')\ ,\\
    &A_m''+\frac{2}{\eta}A_m'-k^2\Psi=\frac{c^{(2)}}{\eta'}\delta(\eta-\eta')\ ,
    \end{split}
    \right.
\end{equation}
where $(c^{(1)},c^{(2)})=(1,0),(0,1)$ and we used that $\ch=2/\eta$ in matter domination. The form of these equations suggests that the Green's functions are actually functions of $\eta/\eta'$ and that they solve the homogeneous problem in \eqref{eq:boltz_einst_y} with initial conditions  in $\eta/\eta'=1$ given by $\Psi^{(i)}(k,1)=\delta_{i1}$, $A_m^{(i)}(k,1)=0$, $A_m^{(i)'}(\vec{k},1)=\delta_{i2}/\eta'$. While there is no closed form for these Green's functions, we provide some asymptotic behaviors below

\begin{equation}\label{eq:psi1}
\begin{split}
    &\Psi^{(1)}\left(k,\frac{\eta}{\eta'}\right)=\left\{
    \begin{aligned}
        &\frac{\eta'^5}{\eta^5}+\frac{\eta'^2k^2}{15},~ &k\eta'\ll 1\\
        &\frac{36}{5}\frac{1}{k^2\eta'^2},~ &k\eta'\gg 1
    \end{aligned}
    \right.,\quad
    A^{(1)}_m\left(k,\frac{\eta}{\eta'}\right)=\left\{
    \begin{aligned}
        &\frac{\eta'^2k^2}{2}\left(\frac{\eta^2k^2}{45}+\frac{2}{3}-\frac{\eta'}{\eta}+\frac{\eta'^3}{3\eta^3}\right),~ & k\eta'\ll 1\\
        &\frac{6}{5}\frac{\eta^2}{\eta'^2},~ &k\eta'\gg 1
    \end{aligned}
    \right.\\
    &\Psi^{(2)}\left(k,\frac{\eta}{\eta'}\right)=\left\{
    \begin{aligned}
        &\frac{1}{5}-\frac{\eta'}{4\eta}+\frac{\eta'^5}{20\eta^5},~ &k\eta'\ll 1\\
        &\frac{6}{5}\frac{1}{\eta'^2k^2}\left(1-\frac{\eta'^5}{\eta^5}\right),~ &k\eta'\gg 1
    \end{aligned}
    \right.,\quad
    A^{(2)}_m\left(k,\frac{\eta}{\eta'}\right)=\left\{
    \begin{aligned}
        &1-\frac{\eta'}{\eta},~ &k\eta'\ll 1\\
        &\frac{1}{5}\left(\frac{\eta^2}{\eta'^2}-\frac{\eta'^3}{\eta^3}\right),~ &k\eta'\gg 1
    \end{aligned}
    \right.\ ,
\end{split}
\end{equation}

and show the full numerical solutions for one subhorizon and one superhorizon mode in Figure \ref{fig:green}.

The full solution then is

\begin{equation}\label{eq:psi_solution}
\begin{split}
    \Psi(\vec{k},\eta)=&~\frac{\Ods}{\beta_H}\left(\frac{12}{k^2\eta_*^2}\hats(\vec{k})\Psi^{(1)}\left(k,\frac{\eta}{\eta_*}\right)-6\hats(\vec{k})\Psi^{(2)}\left(k,\frac{\eta}{\eta_*}\right)+\int_{\eta_*}^{\eta}\frac{\eta_*^2\mathrm{d}\eta'}{\eta'^3}\Psi^{(1)}\left(k,\frac{\eta}{\eta'}\right)\hatr_>(\vec{k},\eta')\right)
    \end{split}
\end{equation}

and an identical expression holds for $A_m(\vec{k},\eta)$. We can thus write the explicit expressions for the scalar transfer functions defined in \eqref{eq:psi}

\begin{equation}\label{eq:scalar_transfer}
\begin{split}
    & T_\sigma(\eta,k)=\frac{12}{k^2\eta_*^2}\Psi^{(1)}\left(k,\frac{\eta}{\eta_*}\right)-6\Psi^{(2)}\left(k,\frac{\eta}{\eta_*}\right)-\frac{\sqrt{3}k\eta_*}{2}\int_{\eta_*}^{\eta}\frac{\eta_*^2\mathrm{d}\eta'}{\eta'^3}\Psi^{(1)}\left(k,\frac{\eta}{\eta'}\right)\sin\left(\frac{k}{\sqrt{3}}(\eta'-\eta_*)\right)\\
    & T_p(\eta,k)=-3\int_{\eta_*}^{\eta}\frac{\eta_*^2\mathrm{d}\eta'}{\eta'^3}\Psi^{(1)}\left(k,\frac{\eta}{\eta'}\right)\left(\frac{k}{\sqrt{3}\ch_*}\sin\left(\frac{k}{\sqrt{3}}(\eta'-\eta_*)\right)+\cos\left(\frac{k}{\sqrt{3}}(\eta'-\eta_*)\right)\right)\ .
\end{split}
\end{equation}

In Figure \ref{fig:transfer} we show the behavior of the different transfer functions with time, both for superhorizion and subhorizon modes.

\section{Parametric suppression of continuous contributions to the ISW}\label{sec:approx_bessel}
In this appendix, we show how the ISW is parametrically dominated by the most discontinuous terms in the transfer functions. We shall use a toy example for the transfer function of the scalar potential $\Psi$, namely

\begin{equation}\label{eq:toy_transfer}
    \Psi(\eta)=c_1\eta_*\delta(\eta-\eta_*)+c_2\theta(\eta-\eta_*)+c_3\left(1-\frac{\eta_*}{\eta}\right)\theta(\eta-\eta_*)\ ,
\end{equation}

so that the expression for the $\mathcal{D}_\ell$ in \eqref{eq:ISWfull} is

\begin{equation}
    \mathcal{D}_\ell\propto\int k^2 \mathrm{d}k f(k) \ \left[c_1k\eta_*j'_\ell(k(\eta_0-\eta_*))+c_2j_\ell(k(\eta_0-\eta_*))+c_3\eta_*\int_{\eta_*}^{\eta_0}\frac{\mathrm{d}\eta}{\eta^2}j_\ell(k(\eta_0-\eta))\right]^2\equiv c_ic_j\mathcal{D}_\ell^{(ij)}\ .
\end{equation}

where we neglected unnecessary constants. We shall also assume $f(k)$ to be flat as $k\rightarrow0$ with a power-law fall-off for $k\gtrsim \beta_*$. To get the parametric estimates of the integrals, we will make use of the following approximations for spherical Bessel functions valid for large $\ell$

\begin{equation}
\begin{split}
    & j_\ell^2(x)\simeq\frac{\pi}{2(2\ell+1)}\delta(x-\ell),\quad \int k^2\mathrm{d}k f(k)j_\ell(k\chi_1)j_\ell(k\chi_2)\simeq \frac{\pi}{2\chi_1^2}f\left(\frac{\ell+\frac{1}{2}}{\chi_1}\right)\delta(\chi_1-\chi_2)\ ,
    \end{split}
\end{equation}

as well as of the fact that $j'^2_\ell(x)$ behaves like $\sin^2(x)/x^2$ for $x>\ell$ and is power-law suppressed for $x<\ell$. We get 

\begin{equation}
    \mathcal{D}_\ell\propto \frac{\beta_*^3\eta_*^2}{(\eta_0-\eta_*)^2}f\left(\frac{\ell}{\eta_0-\eta_*}\right)\left[c_1^2+\frac{c_2^2\ell}{\beta_H^3\left(\frac{\eta_0}{\eta_*}-1\right)}+\frac{c_1c_2\ell}{\beta_H^3\left(\frac{\eta_0}{\eta_*}-1\right)^2}+\frac{(c_1+c_2+c_3)c_3}{\beta_H^3}\right]\ .
\end{equation}

\begin{figure}[htp!]
    \centering
    \includegraphics[width=0.475\linewidth]{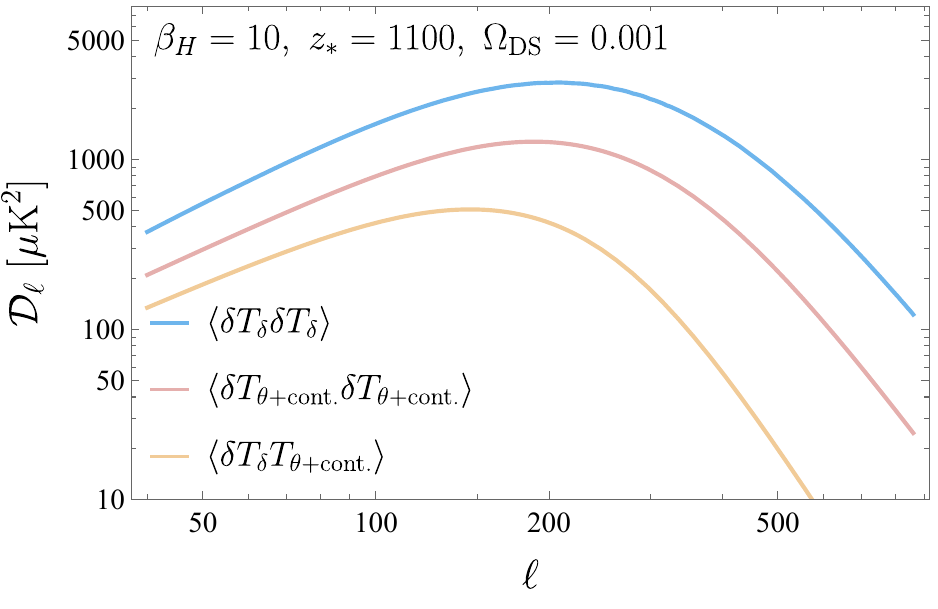}\quad\includegraphics[width=0.475\linewidth]{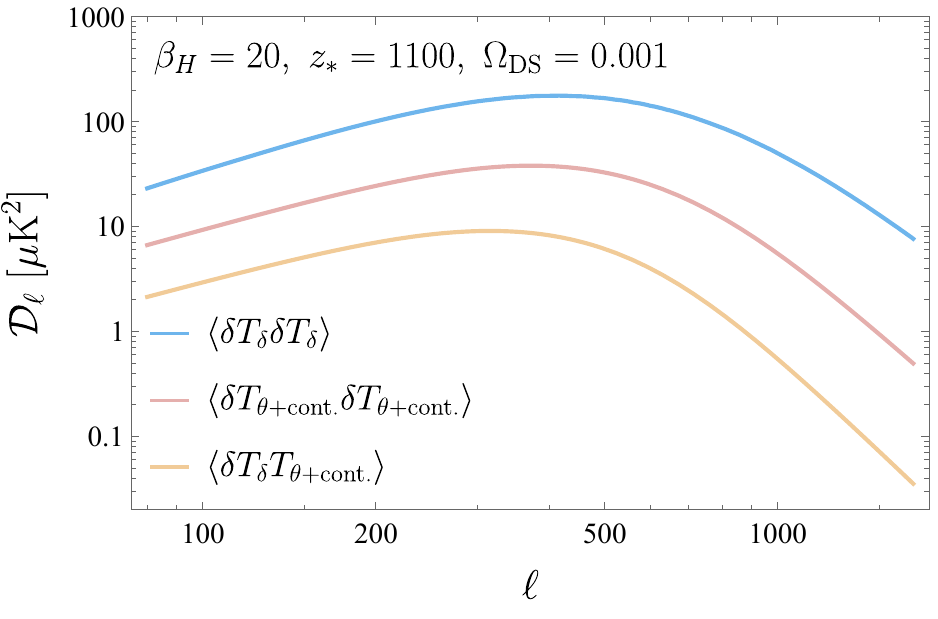}
    \caption{Comparison of the relative contributions to the CMB angular power spectrum arising from the different components of the ISW potential $\Phi-\Psi$, classified according to their temporal discontinuity. Results are shown for $z_*=1100$, $\Ods=10^{-3}$ and $\beta_H=10$ ($\beta_H=10$) in the \textbf{left} (\textbf{right}) panel. The \textbf{blue} curves represent the contribution from the $\delta$-function in $\Phi$, induced by Eq.~\eqref{eq:sigma}; the \textbf{red} curves correspond to the leftover part of the potential, while the \textbf{yellow} curves show their cross-correlation.}
    \label{fig:Dell}
\end{figure}

If we evaluate $\mathcal{D}_\ell$ at the position of the peak, roughly given by $\ell_*\simeq \beta_H\left(\frac{\eta_0}{\eta_*}-1\right)$, we get

\begin{equation}\label{eq:param_estimate}
    \mathcal{D}_\ell\propto \frac{\beta_H^3}{\eta_*(\eta_0-\eta_*)^2}f\left(\frac{\beta_H}{\eta_*}\right)\left[c_1^2+\frac{c_2^2}{\beta_H^2}+\frac{c_1c_2}{\beta_H^2\left(\frac{\eta_0}{\eta_*}-1\right)}+\frac{(c_1+c_2+c_3)c_3}{\beta_H^3}\right]\ ,
\end{equation}

which shows that the contribution from the $c_1\delta(\eta-\eta_*)$ term in \eqref{eq:toy_transfer} is enhanced at least by a factor $\beta_H$ compared to the less discontinuous terms. From \eqref{eq:scalar_decomposition}, the $\Phi-\Psi$ combination that sources the ISW in the post-recombination dynamics considered here can be decomposed into $\delta$, $\theta$ and continuous functions of time as in \eqref{eq:toy_transfer}. Using the complete solution in \eqref{eq:psi_solution}, in Figure \ref{fig:Dell} we explicitly show that also in this case the ISW signal is dominated by the most discontinuous function of time and that, consistently with \eqref{eq:param_estimate}, this approximation becomes more precise as $\beta_H$ grows.

\section{Anisotropic stress spectrum from first order phase transition in the envelope approximation}\label{app:anisotropicstress}
\begin{figure}[t]
    \centering
    \includegraphics[width=0.49\linewidth]{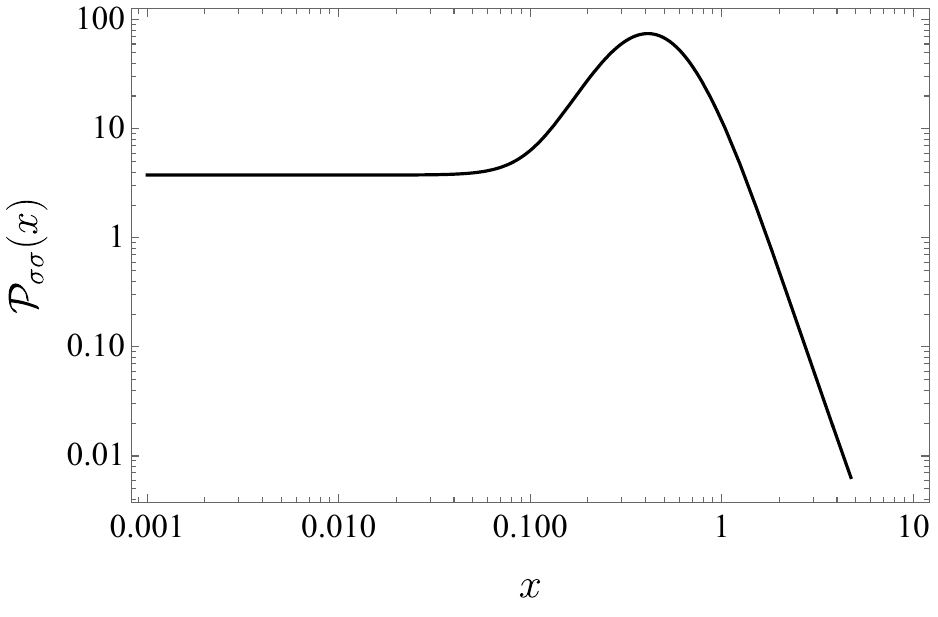}
    \caption{Dimensionless power spectrum of the anisotropic stress, in the dimensionless units defined in Eq.~\eqref{eq:ff}, for a first-order phase transition in the envelope approximation.}
    \label{fig:Pss}
\end{figure}
In this appendix, we compute the anisotropic stress two-point function for first order phase transitions. We recall here again for simplicity the definition of the two-point correlator
\begin{equation}\label{eq:anisotropicstress}
    \langle \hats(\vec{k})\hats(\vec{q})\rangle=\left(\frac{2\pi}{\beta_*}\right)^3\delta(\vec{k}+\vec{q})\mathcal{P}_{\sigma\sigma}(k/\beta_*)\ ,
\end{equation}
where we factored out the dynamical scale $\beta_*$, while the shape of $\mathcal{P}_{\sigma\sigma}(k/\beta_*)$ is a universal function which we aim to compute here. 

 Our analysis is based on the envelope and thin-wall approximations, originally introduced in Ref.~\cite{Kosowsky:1992vn} to estimate the gravitational-wave spectrum from bubble collisions in the simplified configuration of two identical bubbles. A statistical framework to incorporate bubble nucleation effects was subsequently developed in Ref.~\cite{Jinno:2016vai}, enabling the systematic evaluation of correlation functions for tensor perturbations. We adopt this formalism and extend it to the computation of the anisotropic stress spectrum.

In the thin-wall approximation, all the vacuum energy density $\rho_v$ released during bubble expansion is assumed to be localized on the bubble wall. This corresponds to the following form for the surface tension:
\begin{equation}
    \rhov\rho_S(\vec{x}-\vec{x}_n,\eta)=\frac{\rhov}{3}\times\lim_{l_B\rightarrow 0}\left\{
    \begin{split}
        &\frac{r_n(\eta)}{l_B}\ , \ \text{if }r_n(\eta)<|\vec{x}-\vec{x}_n|<r_n(\eta)+l_B\\
        &0\ , \ \text{otherwise}
    \end{split}\right.\, .
\end{equation}

Here $l_B$ denotes the wall thickness, $\vec{x}_n$ the nucleation site, and $r_n(\eta)$ the time-dependent bubble radius. In the following, we assume that bubbles expand at the speed of light, so that for a nucleation time $\eta_n$ one has $r_n(\eta)=\eta-\eta_n$. 

The envelope approximation further neglects the field configurations in the overlap region of colliding bubbles, so that the anisotropic stress is entirely sourced by the ensemble of uncollided bubble walls. Under these assumptions, the Fourier transform of the spatial components of the stress tensor generated by the bubbles can be written as
\begin{equation}\label{eq:stress-tensor}
\begin{aligned}
    T_{ij}(&\vec{k},\eta)=\rhov\sum_n\int_{B_n}\mathrm{d}^3xe^{-i\vec{k}\cdot\vec{x}}\rho_S(\vec{x}-\vec{x}_n,\eta) \hat{n}_i\hat{n}_j\equiv \rhov\int\mathrm{d}^3xe^{-i\vec{k}\cdot\vec{x}}\hat{T}_{ij}(\vec{x},\eta)\ ,
\end{aligned}
\end{equation}
where $\hat{n}_i=(\vec{x}-\vec{x}_n)/(\vert x-x_n\vert)$, the sum above is over all the bubbles and the integral $\int_{B_n}$ is intended only over the uncollided surface of the $n^{\rm th}$ bubble, centered at $\vec{x}_n$. Notice that in \eqref{eq:stress-tensor} we did not include additional trace terms related to the potential energy density of the field configuration responsible for the phase transition, since they cancel out upon projection onto the anisotropic stress $\hat{\sigma}(\vec{k})$
\begin{equation}\label{eq:sigma_def}
    \hat{\sigma}(\vec{k})=\beta_*\frac{3}{2}\left(\hat{k}_i\hat{k}_j-\frac{1}{3}\delta_{ij}\right)\int_{-\infty}^\infty \mathrm{d}\eta'\int\mathrm{d}^3xe^{-i\vec{k}\cdot\vec{x}}\hat{T}_{ij}(\vec{x},  \eta')\ .
\end{equation}
The form factor in Eq.~\eqref{eq:anisotropicstress} can then be written as 
\begin{equation}
    \mathcal{P}_{\sigma\sigma}\left(\kappa\right)=\frac{9}{4}\int_{-\infty}^\infty\mathrm{d}\tau'\int_{-\infty}^\infty\mathrm{d}\tau''\int\mathrm{d}^3\rho e^{-i\vec{\kappa}\cdot\vec{\rho}}\left(\hat{k}_i\hat{k}_j-\frac{1}{3}\delta_{ij}\right)\left(\hat{k}_m\hat{k}_n-\frac{1}{3}\delta_{mn}\right)\langle\hat{T}_{ij}(\vec{x},\tau')\hat{T}_{mn}(\vec{x}+\vec{\rho},\tau'')\rangle\, ,
\end{equation}
where $\vec{\kappa}=\vec{k}/\beta_*$, with all the other space and time coordinates also normalized to $\beta_*$. Like in the case of the gravitational waves considered in \cite{Jinno:2016vai}, the product $\hat{T}_{ij}(\vec{x},\tau')\hat{T}_{mn}(\vec{x}+\vec{\rho},\tau'')$ is non-zero only if the space-time points $(\vec{x},\tau')$ and  $(\vec{x}+\vec{\rho},\tau'')$ are found on a bubble wall. This bubble wall can belong to the same expanding bubble or to two distinct bubbles, so that we also can split $\mathcal{P}_{\sigma\sigma}\left(\kappa\right)$ into a same- and a distinct-bubble contribution (as discussed in Section III of \cite{Jinno:2016vai}),
$\mathcal{P}_{\sigma\sigma}\left(\kappa\right)=\mathcal{P}_{\sigma\sigma}^{(s)}\left(\kappa\right)+\mathcal{P}^{(d)}_{\sigma\sigma}\left(\kappa\right)$. All in all, the calculation closely follows \cite{Jinno:2016vai}, the only difference being that we are not directly computing the spectrum of a metric fluctuation but that of the source stored in the anisotropic stress, so that the projectors on the tensor components of $\delta T_i^j$ are replaced by the projector\footnote{Technically, $\Sigma_{ij}\equiv\hat{k}_i\hat{k}_j-\delta_{ij}/3$ is not a projector, since it does not satisfy $\Sigma_{ij}\Sigma_{jk}=\Sigma_{ik}$.} on the anisotropic stress defined in \eqref{eq:sigma_def}. The final result is
\begin{equation}
    \begin{split}
        & \mathcal{P}_{\sigma\sigma}^{(s)}\left(\kappa\right) = \frac{\pi}{2}\int_0^{\infty}\mathrm{d}\tau_d\int_{\tau_d}^\infty \frac{\rho^2\mathrm{d}\rho}{\mathcal{I}(\tau_d,\rho)}\left(j_0(\kappa\rho)S_0(\rho,\tau_d)+\frac{j_1(\kappa\rho)}{\kappa\rho}S_1(\rho,\tau_d)+\frac{j_2(\kappa\rho)}{(\kappa\rho)^2}S_2(\rho,\tau_d)\right)\\
        & \mathcal{P}_{\sigma\sigma}^{(d)}\left(\kappa\right) = \frac{\pi}{8}\int_0^{\infty}\mathrm{d}\tau_d\int_{\tau_d}^\infty \frac{\rho^2\mathrm{d}\rho}{\mathcal{I}^2(\tau_d,\rho)}\left(j_0(\kappa\rho)-6\frac{j_1(\kappa\rho)}{\kappa\rho}+18\frac{j_2(\kappa\rho)}{(\kappa\rho)^2}\right)\hat{S}_0(\rho,\tau_d)
    \end{split}
\end{equation}
where $\tau_d=\tau'-\tau''$, $j_n(x)$ the $n^{\rm th}$ spherical Bessel function of the first kind, while
\begin{equation}
    \mathcal{I}(\tau_d,\rho)=e^{\tau_d/2}+e^{-\tau_d/2}+\frac{\tau_d^2-(\rho^2+4\rho)}{4\rho}e^{-\rho/2}\ . 
\end{equation}
The function $\mathcal{I}(\tau_d,\rho)$ is related to the probability $f$ of finding both $(\vec{x},\tau')$ and $(\vec{x}+\vec{\rho},\tau'')$ in the false vacuum by $f=\exp\left(8\pi\Gamma\left(\frac{\tau'+\tau''}{2}\right)\mathcal{I}(\tau_d,\rho)\right)$, where $\Gamma(\tau)=\Gamma_* e^{\tau}$ is the bubble nucleation rate per unit volume (also assumed normalized to $\beta_*$). The functions $S_i$ and $\hat{S}_0$ are given by
\begin{equation}
    \begin{split}
        S_0(\rho,\tau_d)=&~\frac{e^{-\rho/2}}{36 \rho^5}
\big(
\rho^4 \left(96 + 48 \rho - 4 \rho^3 + \rho^4 \right)
- 2 \rho^2 \left(288 + 144 \rho + 32 \rho^2 + 4 \rho^3 + \rho^4 \right) \tau_d^2+\\
&+ \left(864 + 432 \rho + 96 \rho^2 + 12 \rho^3 + \rho^4 \right) \tau_d^4
\big)\\
S_1(\rho,\tau_d)=&~-\frac{e^{-\rho/2}}{6 \rho^5}
\big(
\rho^4\left(96  + 48 \rho + 12 \rho^2 + 2 \rho^3 + \rho^4 \right)
- 2\rho^2\left(576  + 288 \rho + 68 \rho^2 + 10 \rho^3 +  \rho^4 \right)\tau_d^2+\\
&
+ \left(1440 + 720 \rho + 156 \rho^2 + 18 \rho^3 + \rho^4 \right)\tau_d^4
\big)\\
S_2(\rho,\tau_d)=&~\frac{e^{-\rho/2}}{2 \rho^5}
\big(
 \rho^4\left(144 + 72 \rho + 20 \rho^2 + 4 \rho^3 + \rho^4 \right)
- 2\rho^2\left(720  + 360 \rho + 84 \rho^2 + 12 \rho^3 +  \rho^4 \right)\tau_d^2+\\
&
+ \left(1680 + 840 \rho + 180 \rho^2 + 20 \rho^3 + \rho^4 \right)\tau_d^4
\big)\\
    \hat{S}_0(\rho,\tau_d)=&~\frac{e^{-\rho} (\rho^2 - \tau_d^2)^2 
\left( \rho^4 (2 + \rho)^2 - (12 + \rho (6 + \rho))^2 \tau_d^2 \right)}
{36 \rho^6}\ .
    \end{split}
\end{equation}

The resulting shape of the anisotropic stress power spectrum is shown in Figure \ref{fig:Pss}. A careful cross-check of this result with full field level simulation is left for a separate study~\cite{toappear}. Notice that while the generic flat shot noise spectrum is recovered on large scales~\cite{Amin:2025dtd} ($k\ll\beta_*$) the presence of a bump for $k\sim\beta_*$ is important to correctly model the ISW contribution.

\end{document}